\documentclass[11pt]{amsart}
\usepackage[T1]{fontenc}
\usepackage{graphicx}
\usepackage[utf8]{inputenc}
\usepackage{url}
\usepackage{hyperref}
\input{xy}
\xyoption{all}
\xyoption{poly}
\usepackage[all]{xy}
\usepackage{float}
\usepackage{enumerate}
\usepackage[usenames]{color}
\usepackage{mathtools}
\usepackage{algorithm}
\usepackage{algpseudocode}
\usepackage{enumitem}
\usepackage{cleveref}
\usepackage{amsopn}
\usepackage{tikz}
\usepackage{xcolor}
\usetikzlibrary{backgrounds, fit, arrows, shapes, calc}
\usepackage{graphbox} 
\usepackage{svg}
\usepackage{subcaption}
\usepackage{pgfplots}

\usepackage{amsmath,amsthm,amsfonts,amssymb}
\theoremstyle{plain}
\newtheorem{theorem}{Theorem}[section]
\newtheorem*{theorem*}{Theorem}

\theoremstyle{definition}

\newtheorem{definition}[theorem]{Definition}

\theoremstyle{remark}

\newcommand{\immath}{\dot{\imath}}
\DeclareMathOperator{\median}{median}

\newcommand{\R}{\mathbb{R}}
\newcommand{\N}{\mathbb{N}}
\newcommand{\Z}{\mathbb{Z}}

\newcommand{\TP}{\mathrm{TP}}
\newcommand{\TN}{\mathrm{TN}}
\newcommand{\FN}{\mathrm{FN}}
\newcommand{\FP}{\mathrm{FP}}
\newcommand{\TOT}{\mathrm{TOT}}

\title{Topological fingerprints for audio identification} 

\author[Reise]{Wojciech Reise$^{\ast}$}\address{$^{\ast}$Equal contributors}
\address{Spotify, EPFL} 
\email{reisewojciech@gmail.com}
\author[Fern\'andez]{Ximena Fern\'andez$^{\ast}$}
\address{Department of Mathematical Sciences, Durham University and Mathematical Institute, University of Oxford} \email{ximena.fernandez@maths.ox.ac.uk}
\author[Dominguez]{Maria Dominguez}
\address{Spotify} 
\email{mdominguez@spotify.com}
\author[Harrington]{Heather A. Harrington}
\address{Mathematical Institute, University of Oxford}
\email{hharrington@maths.ox.ac.uk}
\author[Beguerisse-D\'iaz]{Mariano Beguerisse-D\'iaz} 
\address{Spotify}
\email{marianob@spotify.com}

\begin{document}

\begin{abstract}
We present a topological audio fingerprinting approach for robustly identifying duplicate audio tracks. Our method applies persistent homology on local spectral decompositions of audio signals, using filtered cubical complexes computed from mel-spectrograms. By encoding the audio content in terms of local Betti curves, our topological audio fingerprints enable accurate detection of time-aligned audio matchings. Experimental results demonstrate the accuracy of our algorithm in the detection of tracks with the same audio content, even when subjected to various obfuscations. Our approach outperforms existing methods in scenarios involving topological distortions, such as time stretching and pitch shifting.

\end{abstract}

\subjclass[2010]{
55N31, 
68U10, 
62R40.  
}

\keywords{Content-based Audio Identification, Topological Fingerprints, Persistent Homology,  Signal Processing,
Topological Data Analysis}

\maketitle

\section{Introduction}






 Audio identification is a well-studied problem in digital audio processing. Applications of audio identification range from practical to commercial, such as music recognition from fragment recordings, cover song detection, audio retrieval, and copyright protection  \cite{baluja_audio_2007, mohri_robust_2007,wang_industrial-strength_2003, yan_ke_computer_2005}.
 In practice, the straightforward approach of directly comparing the digitized waveforms is neither efficient nor effective. Therefore, audio signals are typically analyzed using their spectral representation. The complexity of identifying segments of audio arises from the high dimensionality of audio representations, as well as from the significant variability in the representations of perceptually similar recordings.


Audio identification algorithms involve extracting low-dimensional features from  audio tracks and matching them against a database of audio fingerprints. A fingerprint is a summary of an audio signal that encodes information to uniquely identify it. Fingerprints are often extracted from spectral representations of the audio signals in the time-frequency domain, such as spectrograms which are matrices whose columns represent the frequencies present in the signal at a given point in time~\cite{julius_o._smith_spectral_2011}.
Spectrograms are usually visualised as images or heatmaps. Patterns in this image correspond to auditory features that audio ID systems leverage to identify an audio query~\cite{ baluja_audio_2007, julius_o._smith_spectral_2011, wang_industrial-strength_2003, yan_ke_computer_2005}. For example, in the well-known algorithm Shazam \cite{wang_industrial-strength_2003}, the fingerprint of a track is a set of time-frequency triplets that correspond to local high-intensity points in the spectrogram. 
Other approaches take segments of audio (i.e, subsets of columns) from the spectrogram and decompose them using wavelets~\cite{baluja_audio_2007, yan_ke_computer_2005}. 

Audio fingerprinting, when coupled with matching algorithms, aims to robustly identify different versions of the same audio, even in the presence of obfuscations or artefacts. The most popular audio ID algorithms are able 
to recognize obfuscated audio tracks with addition of noise, reverberation and high or low pass filters. However, they encounter challenges with simple obfuscations such as pitch and tempo shifts, as well as more complex distortions with time varying degrees of combined obfuscations~\cite{grosser_music_obfuscator}.

A growing field in computational mathematics, topological data analysis (TDA), develops and employs topological invariants to quantify the multiscale shape and geometry of data \cite{Ghrist08,Carlsson2009TopologyAD,edelsbrunner_topological_2002}. The applicability of TDA for complex and larger datasets relies on stability theorems to handle noisy data \cite{cohen2005stability},
advances in algorithmic implementation \cite{otter_roadmap_2017}, 
as well as its integration in statistical and machine learning frameworks \cite{ali2022survey}. 
An active area of TDA focuses on studying dynamics of data \cite{batko2020conley,cohen2006vines,carlsson2010zigzag,perea2015sliding}, with several applications in signal analysis. 

In this work, we present an audio identification method based on topological descriptors of audio signals. The topological audio fingerprints will enable the robust identification of audio tracks in the presence of a wide range of obfuscation types, including tempo and pitch stretching. Leveraging techniques from TDA, our method utilizes persistent homology \cite{edelsbrunner_topological_2002, Ghrist08, zomorodian_computing_2005} to accurately quantify audio features even in the presence of such audio distortions. As a result, it outperforms the leading method, particularly in complex obfuscation scenarios involving topological-based transformations, such as pitch and tempo shift. We provide strong evidence by correctly identifying challenging examples from an audio ID challenge database of tracks \cite{grosser_music_obfuscator}, demonstrating the potential of topology-based approaches to improve the robustness and reliability of audio identification systems.




\subsection{Contributions} 
We focus on the \textit{$1$ vs $1$} audio identification task, which aims to decide whether two audio tracks correspond to the same audio content. 
We develop a \textit{Topological Audio ID} algorithm consisting of:
\begin{itemize}
    \item A \textit{topological audio fingerprinting method}, which extracts homological features from spectral representations of  audio signals. Our method builds on the theory of persistent homology for image data, based on filtered cubical complexes.
    \item A tailored \textit{comparison method} that robustly identifies similar audio content from the topological audio fingerprints. In particular, we propose a technique for measuring order-preserving properties in time-matchings.
\end{itemize}
We present rigorous tests on 72,000 pairs of songs from the Million Song Dataset~\cite{bertin_million_2011}, modulo a range of obfuscation scenarios generated with the audio editing software {\fontfamily{lmss}\selectfont  SOX}.
Experimentally, our \textit{Topological Audio ID} algorithm proves to have a comparable performance with established methods for \textit{rigid} obfuscation types (such as addition of noise or reverberation), but a superior performance for \textit{topological distortions} such as pitch and tempo shift.
A concrete comparison with the leading method Shazam \cite{wang_industrial-strength_2003} (either the open-source implementation \cite{shazam_open_source} or the version 15.35.0\footnote{This was the current version at the time of writing.} \cite{shazam_2023}) is presented in \cref{tbl:results_by_obfuscation} and \cref{tbl:scores_grosser}. Notably, our topological method has a good performance in the challenge database generated by Ben Grosser using his Music Obfuscator algorithm \cite{grosser_music_obfuscator}. This is set of 8 highly obfuscated tracks that were not detected either by Shazam or the content ID algorithms on YouTube and Soundcloud in 2016, although they are easily recognizable by humans. 

Our results demonstrate the potential of TDA to capture and extract essential features from audio tracks, thereby offering additional tools to improve current audio identification systems.
We conclude with possible pathways for the generalization of our method to the \textit{$1$ vs $N$} problem, which involves searching for similar audio tracks within a database of fingerprints that match a given query sample.


\subsection{Related work}  In recent years, numerous audio identification systems have been developed for copyright protection on media sharing and streaming platforms like YouTube \cite{google_audioID, baluja_audio_2007} and SoundCloud, and also for entertainment. The most widely used audio identification algorithm currently available  is Shazam \cite{wang_industrial-strength_2003}.

Most audio ID systems consists of: a fingerprinting method, a matching algorithm and a searching program \cite{cano2002}. During the last step, a database with the fingerprints of a collection of recordings is created. Then, a query with an unlabeled sample recording is processed to extract its fingerprint, which is subsequently compared with the fingerprints in the database in order to find a matching. 
Shazam  fingerprints are obtained from \textit{constellations} of local maxima in the spectrogram, and the matching algorithm estimates the \textit{alignment} of the fingerprints with the query snippet~\cite{wang_industrial-strength_2003}. The efficiency of the search step relies in the use of combinatorial hash functions for fast indexing of the database of fingerprints.


There are several applications of TDA to music analysis. In \cite{sanderson_computational_2017}, Sanderson et. al.  use Takens' embeddings of waveforms of single musical notes and persistent rank functions to discriminate between musical instruments.
Bendich et. al. represent an audio track as a point cloud, and then use principal component analysis and TDA to create a graph representation of the track which enables analysing its structure \cite{Bendich2018}.
In these cases, a common assumption is that underlying the time-series of the waveform, there is a dynamical system, and its periodic nature is characterized by the persistent homology of its sliding window embedding. 
In \cite{alcalaalvarez2023framework}, the authors also develop a TDA-based fingerprinting method for music data described in  Western notation.

\subsection{Outline}

In \cref{sec:audio_representations}, we review spectral representations of audio signals and here focus on spectrograms. In \cref{sec:topology_spectrograms}, we present how spectrograms can be analysed using persistent homology of cubical complexes. 
To perform audio identification, we propose the topological audio ID algorithm in \cref{sec:fingerprints}, which consists of two parts:  computing a topological audio fingerprint (\cref{sec:fingerprints_spectrogram}) and then matching, scoring and comparing the generated fingerprints (\cref{sec:compare_fingerprints}).
We next present experimental results in \cref{sec:experiments}, which provide empirical evidence of the algorithm's effectiveness and performance. In \cref{sec:discussion}, we discuss the  results and the strengths and limitations of our topological audio ID algorithm.
We review additional algebraic topology basics for image data in \cref{appendix:cubical_complexes}. 

\section{Spectral representations of audio}
\label{sec:audio_representations}

\begin{figure}[htb!] 

\hspace{-70pt}\includegraphics[width = 0.81\textwidth]{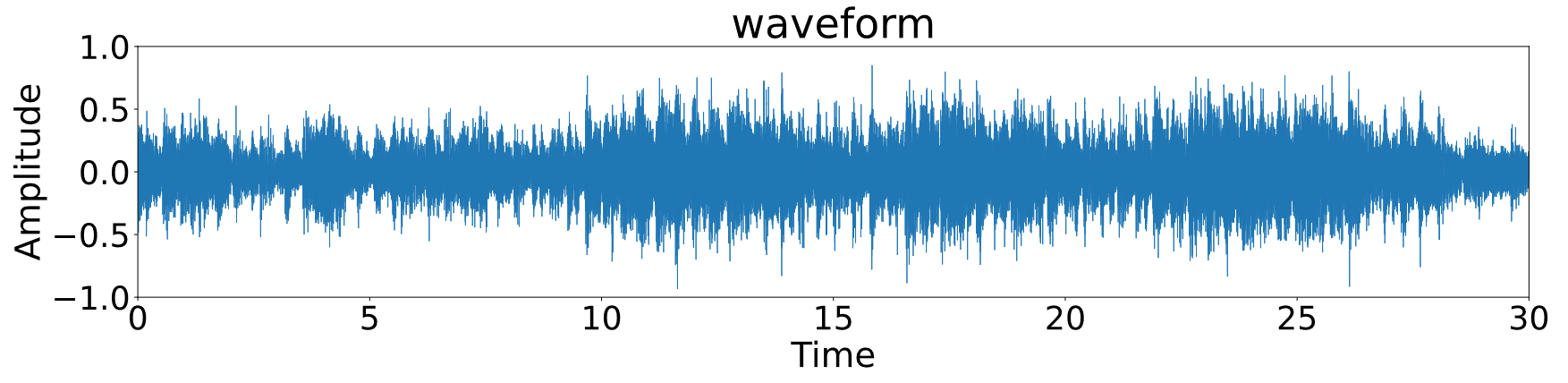}\\
\hspace{10pt}\includegraphics[width = \textwidth]{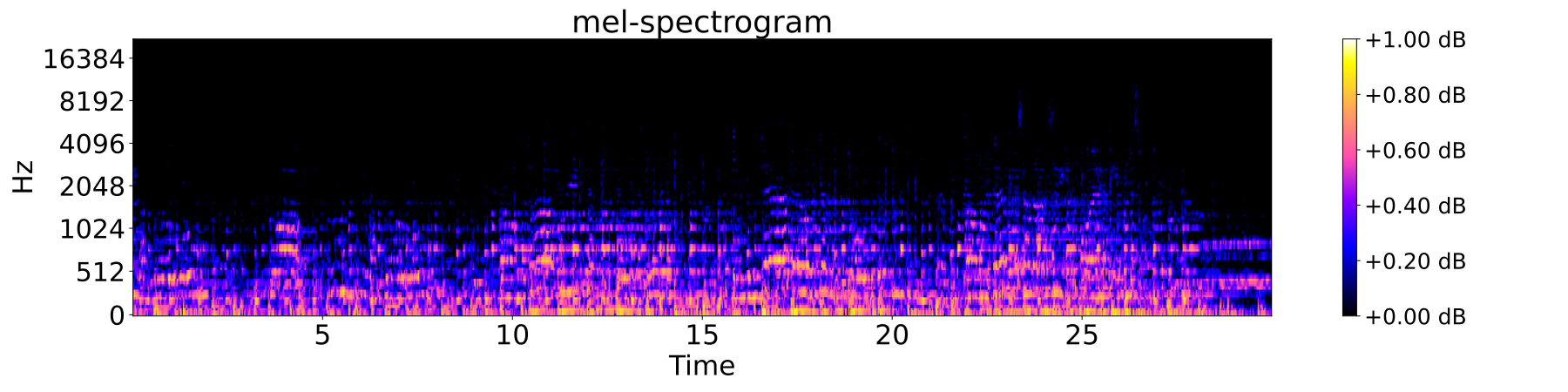}\\
\includegraphics[width = \textwidth, height = 135pt]{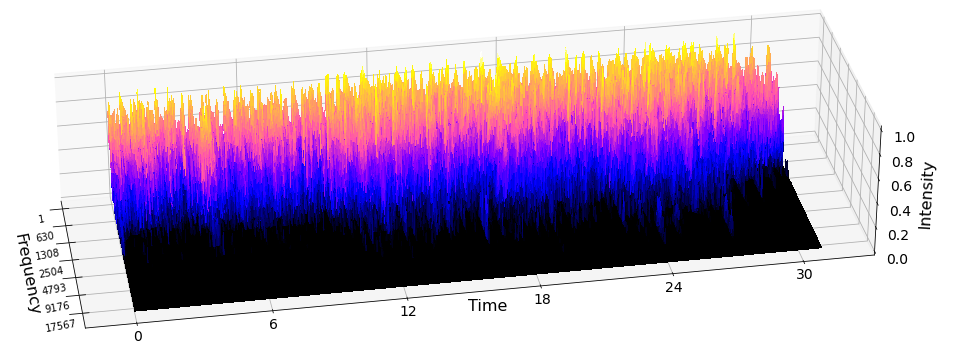}   
\caption{\textbf{Representations of audio.} A 30-second fragment of the track `The Morning' by \textit{Le Loup} \cite{le_loup_morning_2019}. \textit{Top:} The waveform of the audio signal. \textit{Middle:} The \textit{mel-spectrogram} of the waveform. Each column represents the spectral decomposition of the track around a point in time. The lower rows correspond to lower-frequency sounds (e.g., the bass), and the higher rows correspond to high frequency sounds (e.g., violins).
\textit{Bottom:} A 3D representation of the mel-spectrogram, where height is intensity in dB.}
	\label{fig:spectrogram}
 \label{fig:waveform_spectrogram}
\end{figure}

An \textit{audio recording} is the measurement of the pressure fluctuations induced by a \textit{sound wave} in the vicinity of a microphone, which is represented as a continuous real valued function  $s\colon\lbrack 0,T\rbrack \to \R$ over a time interval (i.e., the  \textit{waveform}, see \cref{fig:spectrogram})~\cite{muller2015fundamentals}. 
Although \textit{loudness} of the sound is related to the amplitude of the signal, other audio features such as \textit{pitch} are linked to frequency changes, which cannot be clearly discerned in the waveform representation. The \textit{short-time Fourier transform} (STFT)~\cite{julius_o._smith_spectral_2011} provides a decomposition of the time-varying frequency and phase content of an audio signal. Given $t\in \lbrack 0,T\rbrack$ and a frequency $f\geq 0$, the  \textit{continuous} STFT of the waveform $s$ represents the amplitude of $f$ over a window around $t$. The STFT is computed as 
\begin{equation}
\label{eq:stft}
    S(f,t) = \int_{\R} s(\tau)w(\tau-t) \exp(-\immath f \tau) d\tau,
\end{equation}
where $w(t)$ is a \textit{window function}, which is typically a bell-shaped function centred at zero with finite support~\cite{julius_o._smith_spectral_2011}.

In digital audio processing, the signal is a finite collection of samples $(s_n)_{n=1}^{N}$ of $s$ at equally-spaced time points $(t_n)_{n=1}^{N}$ in $\lbrack 0,T\rbrack$~\cite{muller2015fundamentals}. The size of the sample $N$ is  $T f_s$, where $f_s$ is the \textit{sampling rate}. A spectral representation of $(s_n)_{n=1}^{N}$ can be obtained via the \textit{discrete} STFT~\cite{julius_o._smith_spectral_2011}. Given a discretization of the frequency range $\{f_m\}_{m=1}^M$, the magnitude of the frequency $f_m$ around $t_n$ is
\begin{equation}
\label{eq:discrete_stft}
    \widehat{S}(n, m) = \sum_{k=-\infty}^\infty s_k\,w_{k-n}\exp\left(-\immath k f_m\right),
\end{equation}
where $N_w$ is the size of the window, and $(w_k)_{k=0}^{N_w-1}$ is a discrete version of a window function. In this work, we use the well-known Hann window~\cite{julius_o._smith_spectral_2011}:
\begin{equation}
 w_k =
    \begin{cases}
    \frac{1}{2}\left(1- \cos\left(\frac{2\pi k }{N_w - 1}\right)\right) & \mbox{if } 0 \leq k \leq N_{w},\\
    0 & \mbox{otherwise.}
    \end{cases}
\label{eq:hann_window}
\end{equation}

The \textit{spectrogram} of $(s_n)_{n=1}^{N}$ is a matrix $\mathcal S\in \R^{N\times M}$ whose entries contain the absolute magnitude of the spectral decomposition: 
\begin{equation}
\label{eq:spectrogram}
\mathcal{S}_{n,m} = \left\vert \widehat{S}(n,m) \right\vert.
\end{equation}
The value of $S_{n,m}$ can be interpreted as the \textit{loudness} of a \textit{pitch} with frequency $f_m$ around time $t_n$ in the audio signal.

Because the human ear has a logarithmic frequency resolution rather than linear, the frequency and amplitude scales of spectrograms for audio processing applications use the mel-scale~\cite{stevens_scale_1937} and the Decibel-scale respectively. This spectral representation of an audio signal is called the \textit{mel-spectrogram}.  

A typical visual representation of a spectrogram is as a heat-map or as a 3D surface (Figure~\ref{fig:spectrogram}). In either visualisation, the coordinates are time, frequency and intensity, which can be helpful for interpreting audio features, such as \textit{melody} and \textit{rhythm}.

The relationship between auditory features and their visual patterns in mel-spectrograms can be analyzed using image processing techniques in the problem of audio identification. 
One complication that arises is that the image representation of an audio signal is not invariant under common audio obfuscations (e.g. adding \textit{reverb} or \textit{pitch shifting}). To robustly classify audio tracks under these perturbations, it is usual to associate to each image a local low-dimensional representation known as a \textit{fingerprint}~\cite{baluja_audio_2007, yan_ke_computer_2005}. For example, a fingerprint of an audio signal can be the relative position of local maxima in the spectrogram; these are used in the audio identification algorithm Shazam~\cite{wang_industrial-strength_2003}. Fingerprinting using local maxima is a simple yet powerful representation of audio that allows audio identification under `rigid' obfuscations such as addition of noise or reverb, but it is sensitive to `topological' deformations such as time stretching or pitch-shifting.

We propose a topological fingerprinting technique that is particularly robust to audio obfuscations that represent topological transformations in the image of a spectrogram. We interpret spectrograms geometrically as surfaces in $\R^3$ whose \textit{level sets} display the spatial distribution of different frequencies in time, graded by intensity (see \cref{fig:spectrogram}). Different audio features can be identified as distinct topological patterns in this surface. For example, local maxima, represented as separate connected components in the upper-level subsets, quantify local loud pitches. The `shape' of features in space captures melody and audio patterns up to deformation (see \cref{fig:mel-spectrograms_barcode_betti} in the next section).


\section{Topology of spectrograms}
\label{sec:topology_spectrograms}
We review topological data analysis of image data via cubical complexes and then introduce its application to the study of spectrograms from a geometric point of view. We describe persistent homology applied to analyze the geometric structure of spectrograms and its relationship with the audio features of the underlying signals. 
More mathematical details and definitions are available in \cref{appendix:cubical_complexes}.

\subsection{Combinatorial representation of image data} \label{sec:cubical_ph} 

The topology of image data, such as spectrograms, is best computationally analysed as filtered \textit{cubical complexes} \cite{kaczynski_computational_2011}.
A \textit{cubical complex} $K$ is a collection of cubes, closed under taking faces and intersections (i.e., all faces of a cube from the complex are also in the complex and an intersection of two cubes is also a cube in the complex). 
A \textit{(co)-filtration} of a  cubical complex $K$ is a family  $\{K^i\}_{i=0}^m$ of nested subcomplexes of $K$ satisfying
\[\emptyset = K^m\subseteq K^{m-1}\subseteq \dots \subseteq K^1\subseteq K^0=K.\]

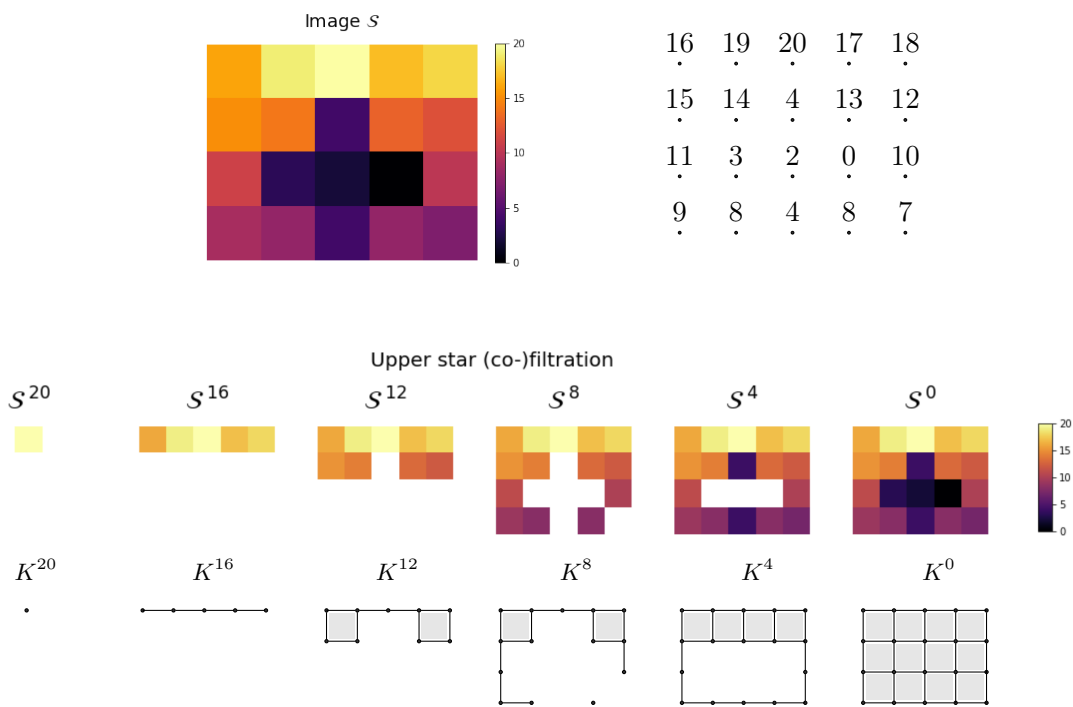
\begin{figure}[htb!]
    \begin{center}
    \includegraphics[width=0.4\textwidth]{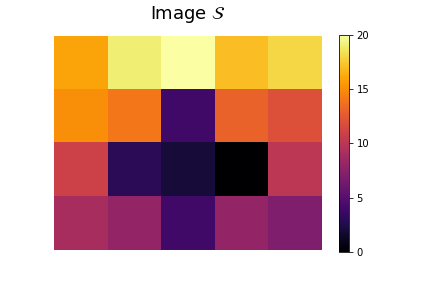}
    \hspace{18pt}
    \begin{tikzpicture}[scale=0.75, general/.style={draw, circle, scale=0.1, fill=black!80, font=\tiny}]

        \node[general, label=above :16] at (0,3) {};
        \node[general, label=above :15] at (0,2) {};
        \node[general, label=above :11] at (0,1) {};
 		\node[general, label=above :9] at (0,0) {};

        \node[general, label=above :19] at (1,3) {};
        \node[general, label=above :14] at (1,2) {};
        \node[general, label=above :3] at (1,1) {};
 		\node[general, label=above :8] at (1,0) {};

        \node[general, label=above :20] at (2,3) {};
        \node[general, label=above :4] at (2,2) {};
        \node[general, label=above :2] at (2,1) {};
 		\node[general, label=above :4] at (2,0) {};

        \node[general, label=above :17] at (3,3) {};
        \node[general, label=above :13] at (3,2) {};
        \node[general, label=above :0] at (3,1) {};
 		\node[general, label=above :8] at (3,0) {};

        \node[general, label=above :18] at (4,3) {};
        \node[general, label=above :12] at (4,2) {};
        \node[general, label=above :10] at (4,1) {};
 		\node[general, label=above :7] at (4,0) {};
        \node at (0,-1){};
 	\end{tikzpicture}


    \vspace{15pt}
    
    \includegraphics[width=\textwidth]{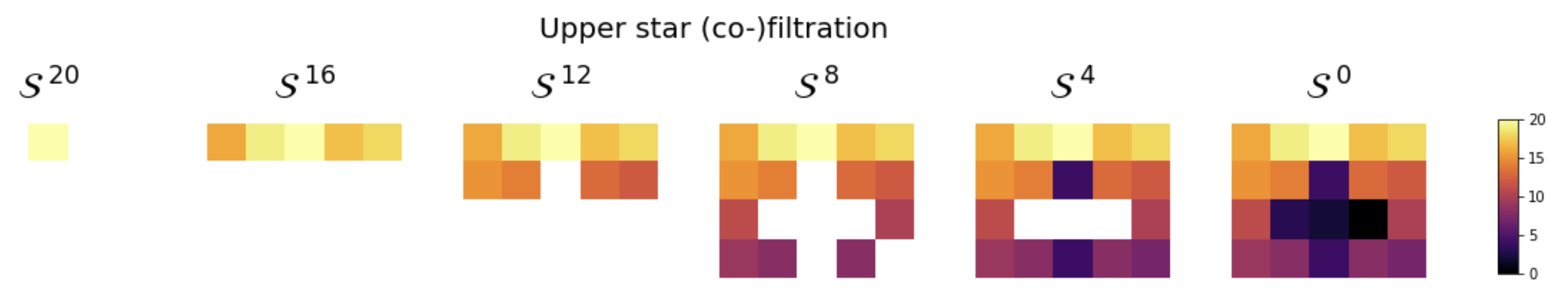}
    \end{center}

    \vspace{20pt}

    \hspace{-55pt}\begin{tikzpicture}[scale=0.41, baseline=-5.5ex, transform shape,
		general/.style={draw,inner sep=0pt,minimum size=3pt},
		dark/.style={circle, fill=black!80},
		arrow/.style={->,>=stealth, shorten >=0.15cm},
		dCurrent/.style={regular polygon, regular polygon sides=3, fill=red!40},
		every label/.append style={font=\tiny}]
		\begin{footnotesize}$K^{20}$\end{footnotesize}
		\node[general, dark] (00) at (-1,-1) {};
		\end{tikzpicture}
        \hspace{35pt}
        \begin{tikzpicture}[scale=0.41, baseline=-5.5ex, transform shape,
		general/.style={draw,inner sep=0pt,minimum size=3pt},
		dark/.style={circle, fill=black!80},
		arrow/.style={->,>=stealth, shorten >=0.15cm},
		dCurrent/.style={regular polygon, regular polygon sides=3, fill=red!40},
		every label/.append style={font=\huge}]
        \begin{footnotesize}$K^{16}$\end{footnotesize}
        \node[general, dark] (-3-1) at (-3,-1) {};
        \node[general, dark] (-2-1) at (-2,-1) {};
        \node[general, dark] (-1-1) at (-1,-1) {};
        \node[general, dark] (1-1) at (1,-1) {};
        \node[general, dark] (0-1) at (0,-1) {};
		\draw  (-3,-1)--(-2-1)--(-1-1)--(0-1)--(1-1);
		\end{tikzpicture}
        \hspace{14pt}
        \begin{tikzpicture}[scale=0.41, baseline=-5.5ex, transform shape,
		general/.style={draw,inner sep=0pt,minimum size=3pt},
		dark/.style={circle, fill=black!80},
		arrow/.style={->,>=stealth, shorten >=0.15cm},
		dCurrent/.style={regular polygon, regular polygon sides=3, fill=red!40},
		every label/.append style={font=\tiny}]
        \begin{footnotesize}$K^{12}$\end{footnotesize}
        \node[general, dark] (-3-2) at (-3,-2) {};
		\node[general, dark] (-2-2) at (-2,-2) {};
        \node[general, dark] (0-2) at (0, -2) {};
        \node[general, dark] (1-2) at (1,-2) {};
        \node[general, dark] (-3-1) at (-3,-1) {};
        \node[general, dark] (-2-1) at (-2,-1) {};
        \node[general, dark] (-1-1) at (-1,-1) {};
        \node[general, dark] (1-1) at (1,-1) {};
        \node[general, dark] (0-1) at (0,-1) {};

        \draw (-3-2) -- (-3-1);
        \draw (-2-2) -- (-2-1);
        \draw (1-2) -- (1-1);
        \draw (0-2) -- (0-1);
        
		\draw (-3-2) -- (-2-2);
        \draw (0-2) -- (1-2);
		\draw  (-3,-1)--(-2-1)--(-1-1)--(0-1)--(1-1);

        \node[fill = gray!20, fit={(0,-1) (1,-2)}, inner sep=-1pt] (A) {};
         \node[fill = gray!20, fit={(-3,-1) (-2,-2)}, inner sep=-1pt] (A) {};
          
		\end{tikzpicture}
        \hspace{14pt}
        \begin{tikzpicture}[scale=0.41, baseline=-5.5ex, transform shape,
		general/.style={draw,inner sep=0pt,minimum size=3pt},
		dark/.style={circle, fill=black!80},
		arrow/.style={->,>=stealth, shorten >=0.15cm},
		dCurrent/.style={regular polygon, regular polygon sides=3, fill=red!40},
		every label/.append style={font=\tiny}]
        \begin{footnotesize}$K^{8}$\end{footnotesize}
		\node[general, dark] (-3-2) at (-3,-2) {};
		\node[general, dark] (-2-2) at (-2,-2) {};
        \node[general, dark] (0-2) at (0, -2) {};
        \node[general, dark] (1-2) at (1,-2) {};
        \node[general, dark] (-3-1) at (-3,-1) {};
        \node[general, dark] (-2-1) at (-2,-1) {};
        \node[general, dark] (-1-1) at (-1,-1) {};
        \node[general, dark] (1-1) at (1,-1) {};
        \node[general, dark] (0-1) at (0,-1) {};
        \node[general, dark] (-3-3) at (-3,-3) {};
        \node[general, dark] (-3-4) at (-3,-4) {};
        \node[general, dark] (1-3) at (1,-3) {};
        \node[general, dark] (0-4) at (0,-4) {};
        \node[general, dark] (-2-4) at (-2,-4) {};

        \draw (-3-4) -- (-3-3) -- (-3-2) -- (-3-1);
        \draw (-2-2) -- (-2-1);
        \draw (1-3) -- (1-2) -- (1-1);
        \draw (0-2) -- (0-1);

        \draw (-3-4) -- (-2-4);
		\draw (-3-2) -- (-2-2);
        \draw (0-2) -- (1-2);
		\draw  (-3,-1)--(-2-1)--(-1-1)--(0-1)--(1-1);

        \node[fill = gray!20, fit={(0,-1) (1,-2)}, inner sep=-1pt] (A) {};
         \node[fill = gray!20, fit={(-3,-1) (-2,-2)}, inner sep=-1pt] (A) {};
		\end{tikzpicture}
        \hspace{13pt}
        \begin{tikzpicture}[scale=0.41, baseline=-5.5ex, transform shape,
		general/.style={draw,inner sep=0pt,minimum size=3pt},
		dark/.style={circle, fill=black!80},
		arrow/.style={->,>=stealth, shorten >=0.15cm},
		dCurrent/.style={regular polygon, regular polygon sides=3, fill=red!40},
		every label/.append style={font=\tiny}]
        \begin{footnotesize}$K^{4}$\end{footnotesize}
		\node[general, dark] (-2-2) at (-2,-2) {};
        \node[general, dark] (-1-2) at (-1,-2) {};
        \node[general, dark] (0-2) at (0, -2) {};
        \node[general, dark] (-1-1) at (-1,-1) {};
        \node[general, dark] (1-3) at (1,-3) {};
        \node[general, dark] (-3-3) at (-3,-3) {};
        \node[general, dark] (-3-2) at (-3,-2) {};
        \node[general, dark] (1-2) at (1,-2) {};
        \node[general, dark] (1-1) at (1,-1) {};
        \node[general, dark] (0-1) at (0,-1) {};
        \node[general, dark] (-2-1) at (-2,-1) {};
        \node[general, dark] (-3-1) at (-3,-1) {};
        \node[general, dark] (-3-4) at (-3,-4) {};
        \node[general, dark] (-2-4) at (-2,-4) {};
        \node[general, dark] (-1-4) at (-1,-4) {};
        \node[general, dark] (0-4) at (0,-4) {};
        \node[general, dark] (1-4) at (1,-4) {};

        \draw (-3-1) -- (-2-1) -- (-1-1) -- (0-1)--(1,-1);
		\draw (-3-2) -- (-2-2) -- (-1-2) -- (0-2)--(1,-2);
        \draw (-3-4) -- (-2-4) -- (-1-4) -- (0-4) --(1,-4);

        \draw (-3-1) -- (-3-2) -- (-3-3) -- (-3-4);
        \draw (-2-1) -- (-2-2);
		\draw (-1-2) -- (-1-1) ;
        \draw (0-1) -- (0-2);
        \draw (1-4) -- (1-3) -- (1-2) -- (1-1);

        \node[fill = gray!20, fit={(-3,-1) (-2,-2)}, inner sep=-1pt] (A) {};
        \node[fill = gray!20, fit={(-2,-1) (-1,-2)}, inner sep=-1pt] (A) {};
        \node[fill = gray!20, fit={(-1,-1) (0,-2)}, inner sep=-1pt] (A) {};
        \node[fill = gray!20, fit=
         {(0,-1) (1,-2)}, inner sep=-1pt] (A) {};
		\end{tikzpicture}
        \hspace{13pt}
        \begin{tikzpicture}[scale=0.41, baseline=-5.5ex, transform shape,
		general/.style={draw,inner sep=0pt,minimum size=3pt},
		dark/.style={circle, fill=black!80},
		arrow/.style={->,>=stealth, shorten >=0.15cm},
		dCurrent/.style={regular polygon, regular polygon sides=3, fill=red!40},
		every label/.append style={font=\tiny}]
  \begin{footnotesize}$K^{0}$\end{footnotesize}
		\node[general, dark] (-2-2) at (-2,-2) {};
        \node[general, dark] (-1-2) at (-1,-2) {};
        \node[general, dark] (0-2) at (0, -2) {};
        \node[general, dark] (-1-1) at (-1,-1) {};
        \node[general, dark] (-2-3) at (-2,-3) {};
        \node[general, dark] (-1-3) at (-1,-3) {};
        \node[general, dark] (0-3) at (0,-3) {};
        \node[general, dark] (1-3) at (1,-3) {};
        \node[general, dark] (-3-3) at (-3,-3) {};
        \node[general, dark] (-3-2) at (-3,-2) {};
        \node[general, dark] (1-2) at (1,-2) {};
        \node[general, dark] (1-1) at (1,-1) {};
        \node[general, dark] (0-1) at (0,-1) {};
        \node[general, dark] (-2-1) at (-2,-1) {};
        \node[general, dark] (-3-1) at (-3,-1) {};
        \node[general, dark] (-2-4) at (-2,-4) {};
        \node[general, dark] (-1-4) at (-1,-4) {};
        \node[general, dark] (0-4) at (0,-4) {};
        \node[general, dark] (1-3) at (1,-4) {};
        \node[general, dark] (-3-4) at (-3,-4) {};

        \draw (-3-1) -- (-2-1) -- (-1-1) -- (0-1) -- (1,-1);
		\draw (-3-2) -- (-2-2) -- (-1-2) -- (0-2) -- (1,-2);
        \draw (-3-3) -- (-2-3) -- (-1-3) -- (0-3) -- (1,-3);
        \draw (-3-4) -- (-2-4) -- (-1-4) -- (0-4) -- (1,-4);

        \draw (-3-1) -- (-3-2) -- (-3-3)-- (-3-4);
        \draw (-2-1) -- (-2-2) -- (-2-3) -- (-2-4);
		\draw (-1-4) -- (-1-3) -- (-1-2) -- (-1-1) ;
        \draw (0-1) -- (0-2) -- (0-3) -- (0,-4);
        \draw (1-4) -- (1-3) -- (1-2) -- (1-1) ;

        \node[fill = gray!20, fit={(-3,-3) (-2,-4)}, inner sep=-1pt] (A) {};
        \node[fill = gray!20, fit={(-2,-3) (-1,-4)}, inner sep=-1pt] (A) {};
        \node[fill = gray!20, fit={(-1,-3) (0,-4)}, inner sep=-1pt] (A) {};
        \node[fill = gray!20, fit={(0,-3) (1,-4)}, inner sep=-1pt] (A) {};
         
        \node[fill = gray!20, fit={(-3,-2) (-2,-3)}, inner sep=-1pt] (A) {};
        \node[fill = gray!20, fit={(-2,-2) (-1,-3)}, inner sep=-1pt] (A) {};
        \node[fill = gray!20, fit={(-1,-2) (0,-3)}, inner sep=-1pt] (A) {};
        \node[fill = gray!20, fit={(0,-2) (1,-3)}, inner sep=-1pt] (A) {};

        \node[fill = gray!20, fit={(-3,-1) (-2,-2)}, inner sep=-1pt] (A) {};
        \node[fill = gray!20, fit={(-2,-1) (-1,-2)}, inner sep=-1pt] (A) {};
        \node[fill = gray!20, fit={(-1,-1) (0,-2)}, inner sep=-1pt] (A) {};
        \node[fill = gray!20, fit={(0,-1) (1,-2)}, inner sep=-1pt] (A) {};
		\end{tikzpicture}
        
    \caption{\textbf{Filtered cubical complex}. \textit{Top:} An image $\mathcal{S}$ representing a small patch of a mel-spectrogram, and the intensity values for every pixel.   \textit{Bottom:} The upper-star co-filtration of the image $\mathcal{S}$ and the cubical vertex construction on $\mathcal{S}$ for some values of the intensity parameter.}
    \label{fig:filtration}
\end{figure}

Given a spectrogram $\mathcal{S}$ and an intensity $i$, we associate a (2-dimensional) cubical complex $K^i$ as follows:
\begin{itemize}
    \item the vertices (or 0-cubes) of $K^i$  are the pixels in the spectrogram whose intensity value is greater than or equal to $i$,
    \item the edges (or 1-cubes) join every pair of vertices in $K^i$  associated to adjacent pixels in $\mathcal S$,
    \item the 2-cubes fill every set of four vertices $\{(m,n), (m+1,n), (m,n+1), (m+1,n+1)\}$ in $K^i$.
\end{itemize}
The construction of $\{K^i\}$ is known as the \textit{upper-star filtration} of the \textit{cubical vertex construction} on $\mathcal{S}$ (see \cref{fig:filtration} for an example). For a complete exposition on the topic we refer the reader to  \cref{appendix:cubical_complexes}.

\subsection{Persistent homology}
\label{sec:persistent_homology}
Homology is a computable topological invariant \cite{Hatcher02}.
Computing homology groups of cubical complexes encodes geometric features of image data at different dimensions. For instance, homology at degree 0, denoted as $H_0$, encodes the number of connected components, whereas $H_1$ reflects the number of \textit{independent} 1-dimensional holes (see \cite{Hatcher02}). 
The rank of the homology groups in dimension $k$ is called the $k$-\textit{Betti number}, which is the number of linearly independent generators of $H_k$. The $0$th-\textit{Betti number} gives the number of connected components whereas the $1$th-\textit{Betti number} gives the number of 1-dimensional holes or loops.
For mathematical definitions of homology, see \cref{appendix:homology}.

 \begin{figure}[htb!]
 	\centering

        \begin{tikzpicture}[scale=0.6, baseline=2ex, transform shape,
		general/.style={draw,inner sep=0pt,minimum size=3pt},
		dark/.style={circle, fill=black!80},
		arrow/.style={->,>=stealth, shorten >=0.15cm},
		dCurrent/.style={regular polygon, regular polygon sides=3, fill=red!40},
		every label/.append style={font=\large}]
        \begin{small}$K^{8}$\end{small}
		\node[general, dark, label=left:$F$] (-3-2) at (-3,-2) {};
		\node[general, dark, label=below :$G$] (-2-2) at (-2,-2) {};
        \node[general, dark, label=below :$I$] (0-2) at (0, -2) {};
        \node[general, dark, label= right :$J$] (1-2) at (1,-2) {};
        \node[general, dark, label=above left:$A$] (-3-1) at (-3,-1) {};
        \node[general, dark, label=above:$B$] (-2-1) at (-2,-1) {};
        \node[general, dark, label=above:$C$] (-1-1) at (-1,-1) {};
        \node[general, dark, label=above right :$E$] (1-1) at (1,-1) {};
        \node[general, dark, label=above:$D$] (0-1) at (0,-1) {};
        \node[general, dark, label=left:$K$] (-3-3) at (-3,-3) {};
        \node[general, dark, label=below left:$O$] (-3-4) at (-3,-4) {};
        \node[general, dark, label=right:$L$] (1-3) at (1,-3) {};
        \node[general, dark, label=below:$R$] (0-4) at (0,-4) {};
        \node[general, dark, label=below:$P$] (-2-4) at (-2,-4) {};

        \draw (-3-4) -- (-3-3) -- (-3-2) -- (-3-1);
        \draw (-2-2) -- (-2-1);
        \draw (1-3) -- (1-2) -- (1-1);
        \draw (0-2) -- (0-1);

        \draw (-3-4) -- (-2-4);
		\draw (-3-2) -- (-2-2);
        \draw (0-2) -- (1-2);
		\draw  (-3,-1)--(-2-1)--(-1-1)--(0-1)--(1-1);

        \node[fill = gray!20, fit={(0,-1) (1,-2)}, inner sep=-1pt] (A) {};
         \node[fill = gray!20, fit={(-3,-1) (-2,-2)}, inner sep=-1pt] (A) {};
		\end{tikzpicture}
        \hspace{30pt}
        \begin{tikzpicture}[scale=0.6, baseline=2ex, transform shape,
		general/.style={draw,inner sep=0pt,minimum size=3pt},
		dark/.style={circle, fill=black!80},
		arrow/.style={->,>=stealth, shorten >=0.15cm},
		dCurrent/.style={regular polygon, regular polygon sides=3, fill=red!40},
		every label/.append style={font=\large}]
        \begin{footnotesize}$K^{4}$\end{footnotesize}
        \node[general, dark] (-1-2) at (-1,-2) {};
        \node[general, dark] (1-3) at (1,-3) {};
        \node[general, dark] (1-1) at (1,-1) {};
        \node[general, dark] (0-1) at (0,-1) {};
        \node[general, dark] (-2-1) at (-2,-1) {};
        \node[general, dark] (-3-1) at (-3,-1) {};
        \node[general, dark] (-3-4) at (-3,-4) {};
        \node[general, dark] (-1-4) at (-1,-4) {};
        \node[general, dark] (0-4) at (0,-4) {};
        \node[general, dark] (1-4) at (1,-4) {};

        \node[general, dark, label=left:$F$] (-3-2) at (-3,-2) {};
		\node[general, dark, label=below :$G$] (-2-2) at (-2,-2) {};
  		\node[general, dark, label=below :$H$] (-1-2) at (-1,-2) {};
        \node[general, dark, label=below :$I$] (0-2) at (0, -2) {};
        \node[general, dark, label= right :$J$] (1-2) at (1,-2) {};
        \node[general, dark, label=above left:$A$] (-3-1) at (-3,-1) {};
        \node[general, dark, label=above:$B$] (-2-1) at (-2,-1) {};
        \node[general, dark, label=above:$C$] (-1-1) at (-1,-1) {};
        \node[general, dark, label=above right :$E$] (1-1) at (1,-1) {};
        \node[general, dark, label=above:$D$] (0-1) at (0,-1) {};
        \node[general, dark, label=left:$K$] (-3-3) at (-3,-3) {};
        \node[general, dark, label=below left:$O$] (-3-4) at (-3,-4) {};
        \node[general, dark, label=right:$L$] (1-3) at (1,-3) {};
        \node[general, dark, label=below:$R$] (0-4) at (0,-4) {};
        \node[general, dark, label=below right:$S$] (1-4) at (1,-4) {};
        \node[general, dark, label=below:$P$] (-2-4) at (-2,-4) {};
        \node[general, dark, label=below:$Q$] (-1-4) at (-1,-4) {};

        \draw (-3-1) -- (-2-1) -- (-1-1) -- (0-1)--(1,-1);
		\draw (-3-2) -- (-2-2) -- (-1-2) -- (0-2)--(1,-2);
        \draw (-3-4) -- (-2-4) -- (-1-4) -- (0-4) --(1,-4);

        \draw (-3-1) -- (-3-2) -- (-3-3) -- (-3-4);
        \draw (-2-1) -- (-2-2);
		\draw (-1-2) -- (-1-1) ;
        \draw (0-1) -- (0-2);
        \draw (1-4) -- (1-3) -- (1-2) -- (1-1);

        \node[fill = gray!20, fit={(-3,-1) (-2,-2)}, inner sep=-1pt] (A) {};
        \node[fill = gray!20, fit={(-2,-1) (-1,-2)}, inner sep=-1pt] (A) {};
        \node[fill = gray!20, fit={(-1,-1) (0,-2)}, inner sep=-1pt] (A) {};
        \node[fill = gray!20, fit=
         {(0,-1) (1,-2)}, inner sep=-1pt] (A) {};
		\end{tikzpicture}
  
 		\caption{\textbf{Homology of a cubical complex.} The cubical complexes $K^8$ and $K^4$ from \cref{fig:filtration}.
   $K^8$ has a two connected components and no 1-dimensional cycles, so $H_0(K^8)$ has a two generators and $H_1(K^8)$ is trivial. 
   On the other hand, $H_1(K^4)$ has rank 1. 
   }
 	\label{fig:homology_cubical_complex}
 \end{figure}
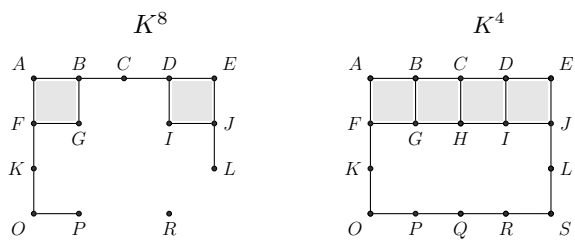

Persistent homology \cite{edelsbrunner_topological_2002, Ghrist08, zomorodian_computing_2005,Carlsson2009TopologyAD} is an extension of classic homology theory to parameterized families of spaces, aiming to capture the evolution of homology groups of (co-)\textit{filtered complexes}. 
Let $(K^i)_{i\in\mathcal{I}}$ be a co-filtration  of complexes, for  $I\subset \N$ finite.
The inclusion maps $K^i \hookrightarrow K^j$ for $i\geq j$ induce maps between the homology groups $\iota_{i,j}: H_k(K^i) \rightarrow H_k(K^j)$. 
We are particularly interested in the cases when $\iota_{i,j}$ is not a bijection, as this implies either the creation or annihilation of a non-trivial homology class.
Then, for a non-surjective (or non-injective) map $\iota_{i, i-1}$ we say that a homology class is born (resp. dies) at $K^{i-1}$. We call $i-1$ a birth (resp. death) value.
The \textit{persistent homology} of a filtered complex is the multi-set  of birth-death pairs $(b,d)$ associated to generators in homology of every dimension. Note that, since $(K^i)_{i\in\mathcal{I}}$ is a co-filtration, $b>d$. A popular graphical representation of persistent homology is the \textit{persistence barcode}, a collection of horizontal line segments in a plane encoding the birth-death pairs $(b,d)$, where the horizontal axis corresponds to the parameter $i$ and the vertical axis represents an (arbitrary) ordering of homology generators. 

\begin{figure}[htb!]
	\centering
    \includegraphics[width = 0.47 \textwidth]{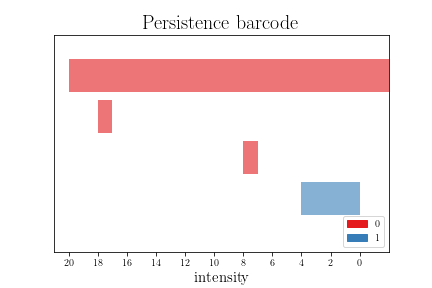} 
    \hspace{2pt}
	\begin{tikzpicture}
	\begin{axis}[axis lines=middle,
	xmin=-2, xmax=22,
	ymin=0, ymax=2.6,
	width=6.4cm, height=4.9cm,
	extra x ticks={0},
    xlabel = {intensity},
    xlabel style={at={(current axis.left of origin)},anchor=north east},
    ylabel = {Betti},
    x dir = reverse, 
    x tick label style = {font = \scriptsize},
    label style={font=\scriptsize},
    legend style={at={(-0.25,0.7)},
    anchor=south west, font=\fontsize{5}{6}\selectfont},
    every y tick label/.style={
    anchor=near yticklabel opposite,
    xshift=0.2em, yshift = +0.5em, font = \scriptsize}]
    \addplot[line width=1.5pt, domain=7:-2, red] {1};
	\addplot[line width=1.5pt, domain=8:7, red, forget plot] {2};
	\addplot[line width=1.5pt, domain=17:8, red, , forget plot] {1};
	\addplot[line width=1.5pt, domain=18:17, red , forget plot] {2};
	\addplot[line width=1.5pt, domain=20:18, red, forget plot] {1};
	\addlegendentry{$\beta_0$};
	\addplot[line width=1.5pt, domain=20:4, blue]{0};
    \addplot[line width=0.7pt, domain=4:0, blue]{1};
    \addplot[line width=1.5pt, domain=0:-2, blue]{0};
	\addlegendentry{$\beta_1$};
    \addplot[color=red,only marks,mark=*] coordinates{(17,1)(8,2) (7,1)(18,2)};
	\addplot[color=red,fill=white,only marks,mark=*] coordinates{(17,2)(7,2)(8,1)(18,1)};
    \addplot[color=blue,only marks,mark=*] coordinates{(4,1)(0,0)};
    \addplot[color=blue,fill=white, only marks,mark=*] coordinates{(4,0)(0,1)};
	\draw[dotted, color = red] (axis cs:8,1) -- (axis cs:8,2);
	\draw[dotted, color = red] (axis cs:7,1) -- (axis cs:7,2);
	\draw[dotted, color = red] (axis cs:18,2) -- (axis cs:18,1);
    \draw[dotted, color = red] (axis cs:17,2) -- (axis cs:17,1);
    \draw[dotted, color = blue] (axis cs:4,0) -- (axis cs:4,1);
	\end{axis}
	\end{tikzpicture}
	\caption{\textbf{Persistent homology} of the (co)-filtered cubical complex of \cref{fig:filtration}. We represent the horizontal axis with opposite orientation, from $+\infty$ to $-\infty$. \textit{Left: } Persistence barcode for dimension 0 and 1. \textit{Right:} The Betti curves $\beta_0$ and $\beta_1$.
	}
 \label{fig:barcode_betti_curve}
\end{figure}
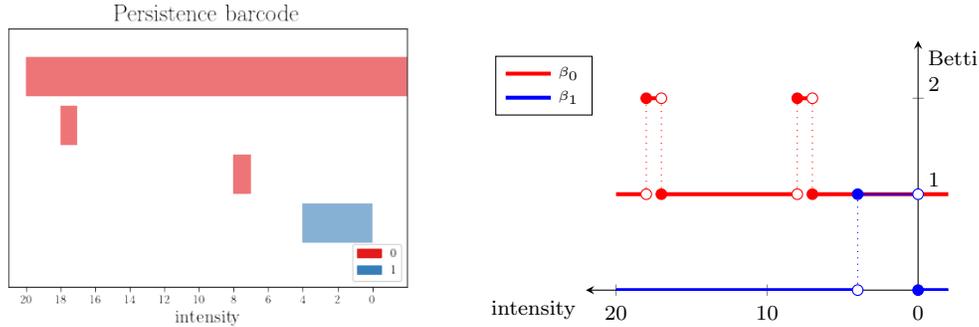

For a filtration, we often visualize the barcode with bars $[b,d)$; in the case of a co-filtration the parameter index deacreases along the horizontal.
An example of a persistence barcode $B$ for persistent homology of dimensions 0 and 1 is shown in \cref{fig:barcode_betti_curve}. 

A functional signature of a barcode $B_k$ of dimension $k$ is given by its \textit{Betti curve} $\beta_k$ \cite{sizemore_classification_2016}, which encode the evolution of the Betti numbers throughout the filtration as
\begin{equation*}
\begin{array}{cccc}
\beta_k:& \R &\rightarrow &\N \\
&x&\mapsto& \sum\limits_{(b, d)\in B_k} 1_{(d, b]}(x).
\end{array}
\label{eq:betti_curve}
\end{equation*}
Here $1_{(d, b]}(x) = 1$  if $d < x \leq b$ and 0 otherwise. We remark that this is a modified definition of Betti curves to take into account that birth is greater than death in a co-filtration. See \cref{fig:barcode_betti_curve} for an example.

For a complete introduction to persistent homology we refer the reader to \cite{otter_roadmap_2017, goodman_persistent_2008, kaczynski_computational_2011}, whereas for algorithmic details on the computation of persistent homology of cubical data see \cite{wagner_efficient_2012}.

\subsection{Topological signatures of spectrograms and audio features} We next propose the geometric analysis of audio tracks based on topological features of spectrograms. Given a spectrogram $\mathcal{S}$, we compute the persistent homology of the upper star co-filtration of the cubical construction on $\mathcal{S}$. 
Different audio features, such as melodies at specific intensities, can be interpreted in the topological fingerprints, such as peaks in the Betti curves.
The lengths of the bars in the persistence barcodes are linked to the (intensity's) depth of the cycles in the surface representation of the spectrogram. 

\begin{figure}[H]
\centering
\includegraphics[width = 0.97\textwidth]
{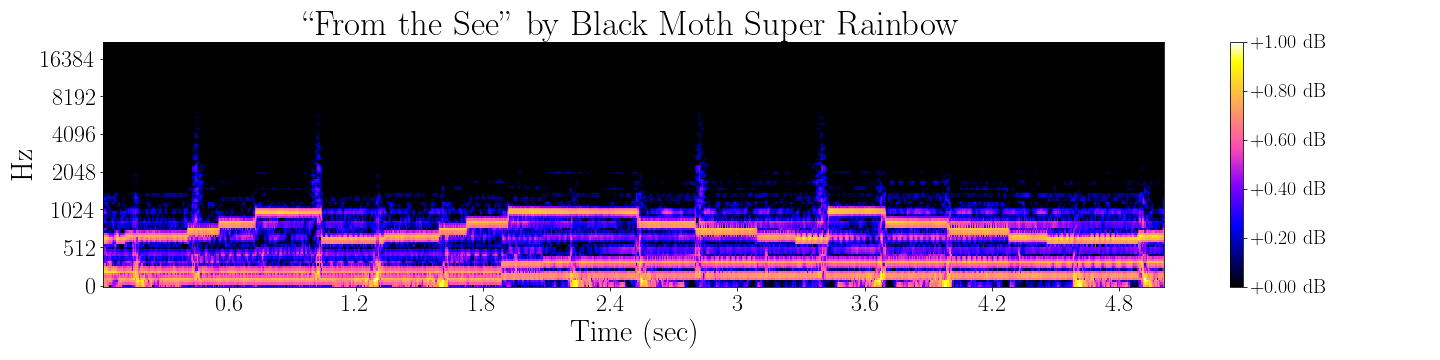}

\vspace{2pt}
 \includegraphics[width = 0.244\textwidth]{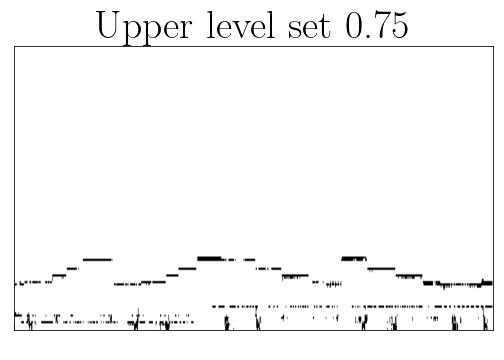}
 \includegraphics[width = 0.244\textwidth]{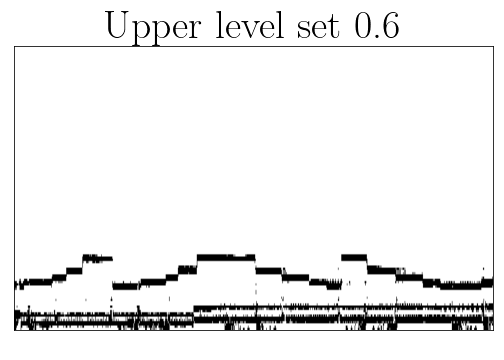}
\includegraphics[width = 0.244\textwidth]{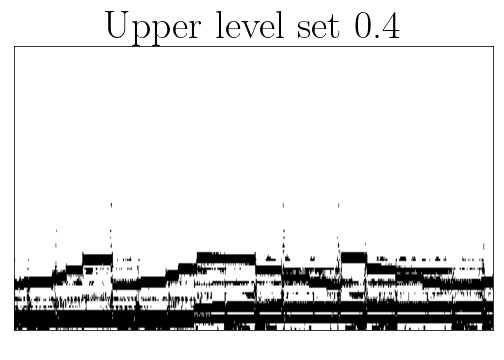}
\includegraphics[width = 0.244\textwidth]{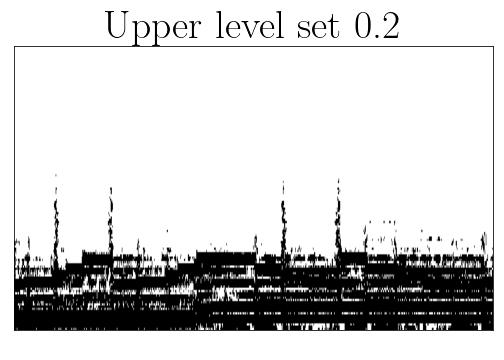}

\includegraphics[width = 0.244\textwidth]{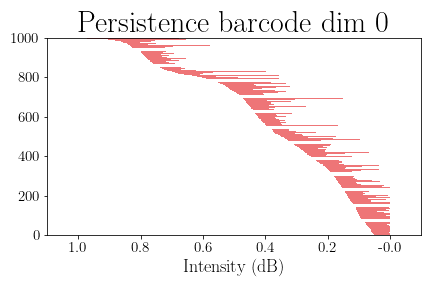} \includegraphics[width = 0.244\textwidth]{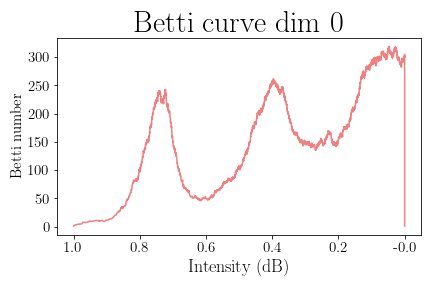}
\includegraphics[width = 0.244\textwidth]{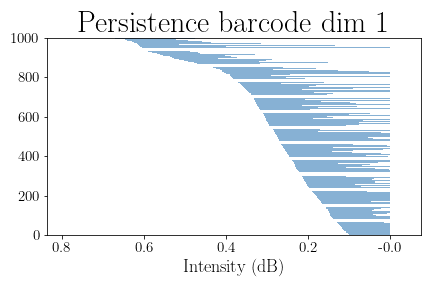} \includegraphics[width = 0.244\textwidth]{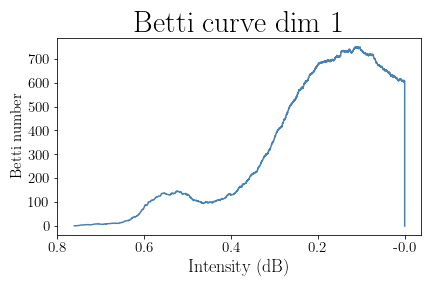}

\vspace{10pt}

\includegraphics[width = 0.97\textwidth]
{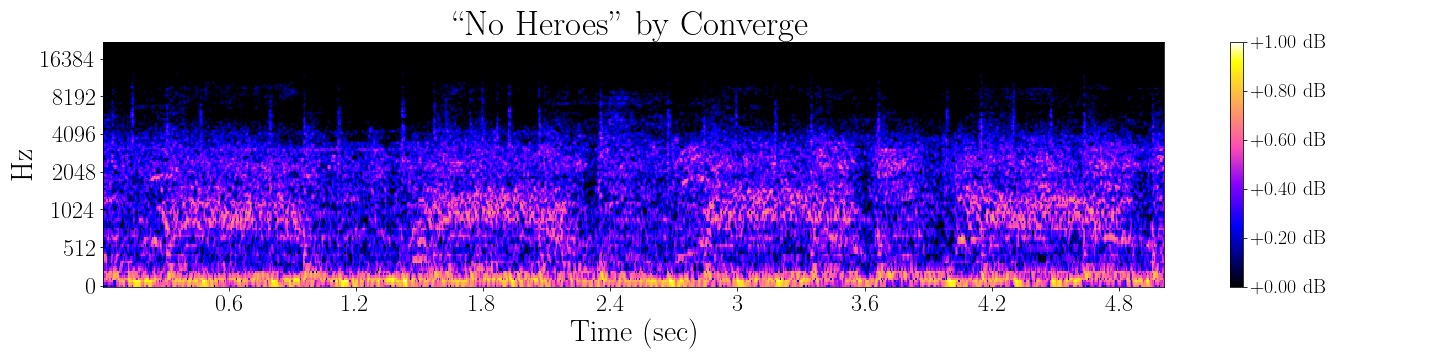}

 \includegraphics[width = 0.244\textwidth]{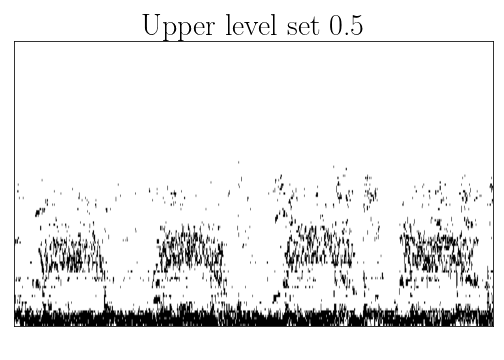}
 \includegraphics[width = 0.244\textwidth]{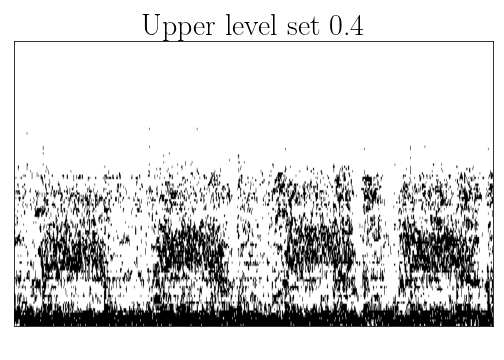}
\includegraphics[width = 0.244\textwidth]{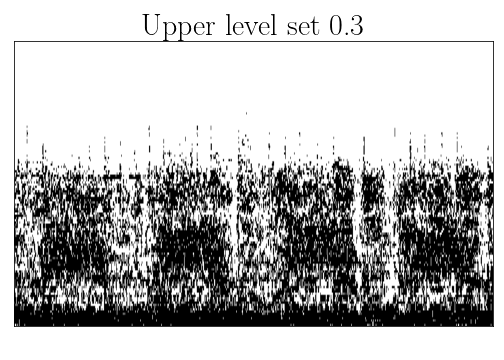}
\includegraphics[width = 0.244\textwidth]{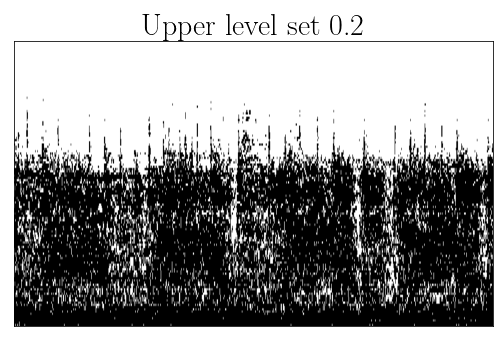}

\includegraphics[width = 0.244\textwidth]{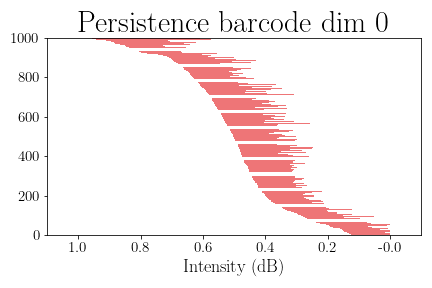} \includegraphics[width = 0.244\textwidth]{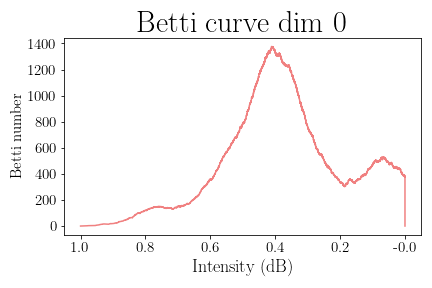}
\includegraphics[width = 0.244\textwidth]{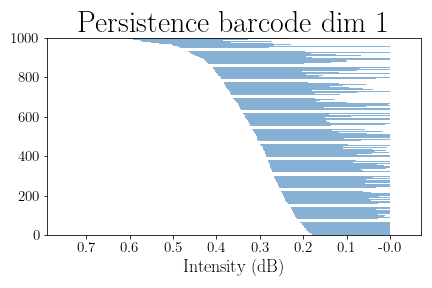} \includegraphics[width = 0.244\textwidth]{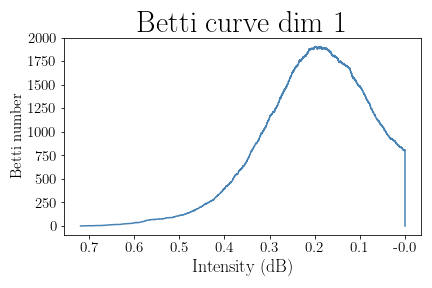}

\caption{\textbf{Topology of spectrograms.}  Mel-spectrogram and its associated persistence barcodes, Betti curves (dimensions 0 and 1) for a 5-seconds extract of different types of audio tracks and a subset of the upper level filtration. \textit{Top:}  Song `From the See' by the psychedelic  indie rock band \textit{Black Moth Super Rainbow}, with a clean electronic melody. 
\textit{Bottom:} Song `No Heroes' from the metalcore band \textit{Converge}, with a strong heavy metal rhythm.
}
\label{fig:mel-spectrograms_barcode_betti}
\end{figure}

In \cref{fig:mel-spectrograms_barcode_betti}, we provide an example of two 5-seconds extracts of different tracks and how their marked audio styles are displayed on their associated barcodes and Betti curves. The first extract corresponds to the song `From the See' by the psychedelic indie rock band \textit{Black Moth Super Rainbow}. It has a clean electronic melody organized around  0.75 and 0.4 decibels, as quantified in the two peaks of its 0-dimensional Betti curve. The 1-dimensional persistence barcode gives a few long bars, which corresponds to the high intensity, large 1-dimensional cycles in the spectrogram. The second extract belongs to the song `No Heroes' from the metalcore band \textit{Converge}. 
The strong heavy metal rhythm in `No Heroes' is observed in the mel-spectrogram \cref{fig:mel-spectrograms_barcode_betti} by many disorganized local maxima at the intensity level $\sim$0.4 dB. Rather than visualising the mel-spectrogram, this audio feature is quantified and identified by a peak in the 0-dimensional Betti curve, exactly at the level $\sim$0.4 dB. At a lower intensity ($\sim$0.2 dB), this topological approach quantifies 3000 long intervals in the one-dimensional barcode, which describes 3000 small cycles in the audio track (see the upper level filtration in \cref{fig:mel-spectrograms_barcode_betti} bottom). 



There is a range of audio obfuscation techniques that allow to modify an audio signal without affecting its recognition by human ear. 
The most popular obfuscation methods are: 
\begin{enumerate}
    \item \textit{Noise:} Addition of a random noise to the main signal. The most common types of noise are: \textit{white noise}, with constant power spectral density (implying that all frequency ranges are affected equally), and  \textit{pink noise}, whose power spectral density logarithmically decreases as frequency increases.
    \item \textit{Reverb:} Addition of reverberation to the main signal. Reverberation is the persistence of the sound after the original sound has stopped and it is generated using the algorithm \textit{freeverb} \cite{shroeder_audio_1961, Schroeder1962NaturalSA}.
    \item \textit{High-pass filter:} A filter that allows only the higher-frequency content of the signal to pass through. This is accomplished by specifying a cutoff frequency, below which signals are attenuated, and above which signals are allowed to pass through with little or no attenuation. It is reflected on the darkening or reduction of intensity in the lower frequency region of the spectrogram.
    \item \textit{Low-pass filter:} Analog to high-pass filter, but removing high-frequency components from the signal and letting only the low-frequency content pass through. It  darkens the upper frequency region of the spectrogram.
    \item \textit{Tempo shift:} Time stretch the audio signal without changing its pitch. This effect uses WSOLA algorithm \cite{wsola_1993}. 
    It is reflected as a continuous linear deformation in time of the spectrogram.
    \item \textit{Pitch shift:} Alter the frequencies of signal components while preserving their harmonic relationships. 
    This transformation is evidenced as a continuous vertical deformation in the spectrogram.
\end{enumerate}
We classify transformations $1-4$ as \textit{rigid} obfuscations, whereas transformations $5-6$ are referred to as \textit{topological} obfuscations. Note that, unlike  rigid obfuscations, topological obfuscations involve the \textit{distortion} of time or frequency variables, resulting in a continuous deformation of the spectrogram either in the horizontal or vertical direction.

\begin{figure}[H]
\includegraphics[width=0.71\textwidth]{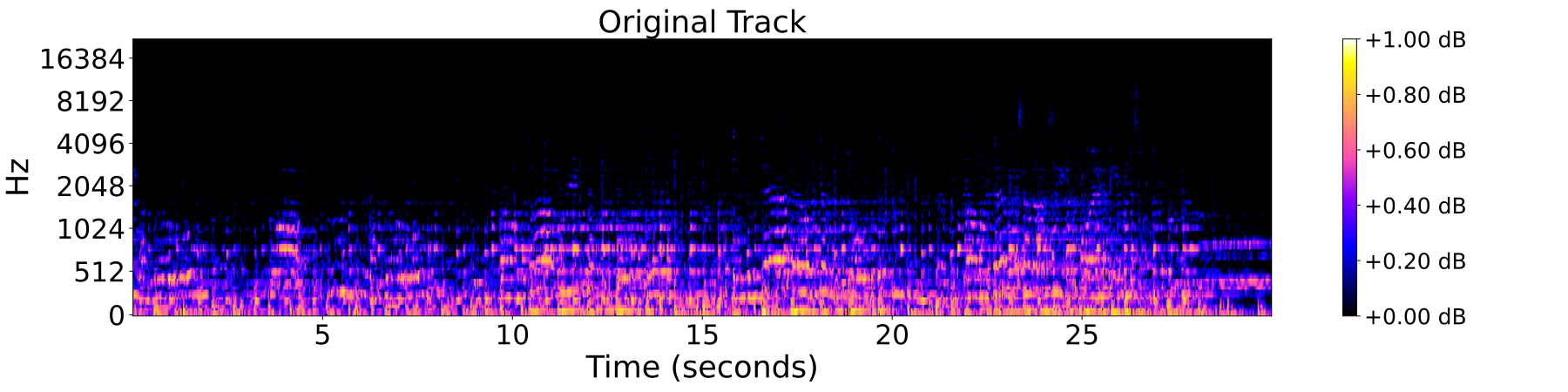} 
\includegraphics[width=0.25\textwidth]{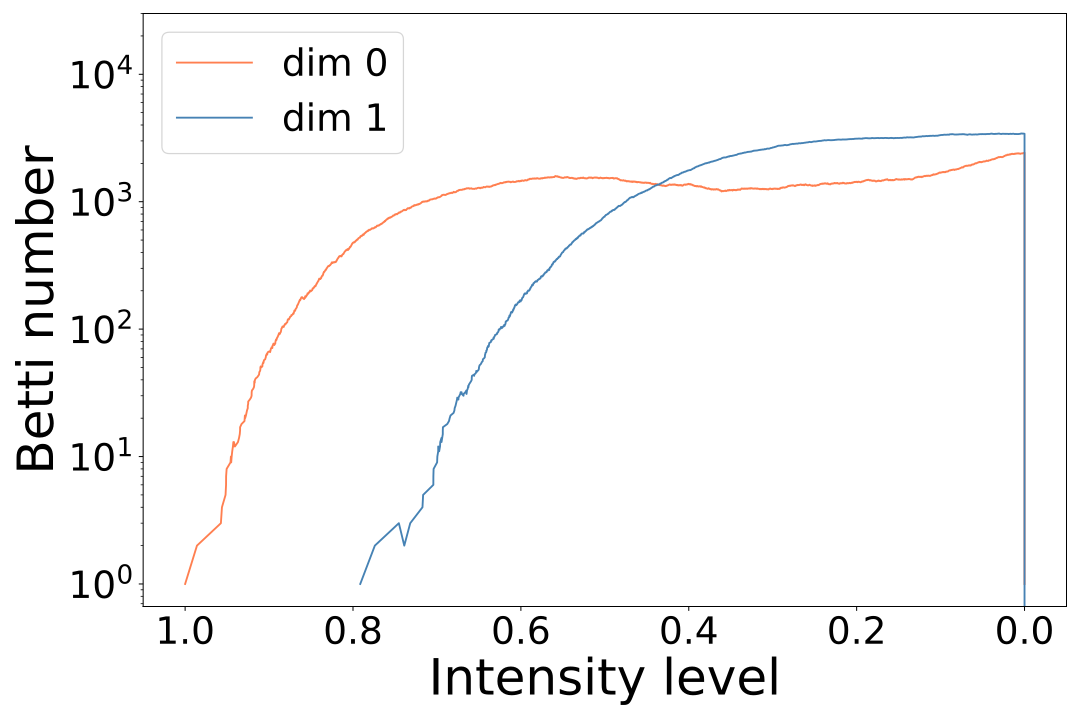}

\includegraphics[width=0.71\textwidth]{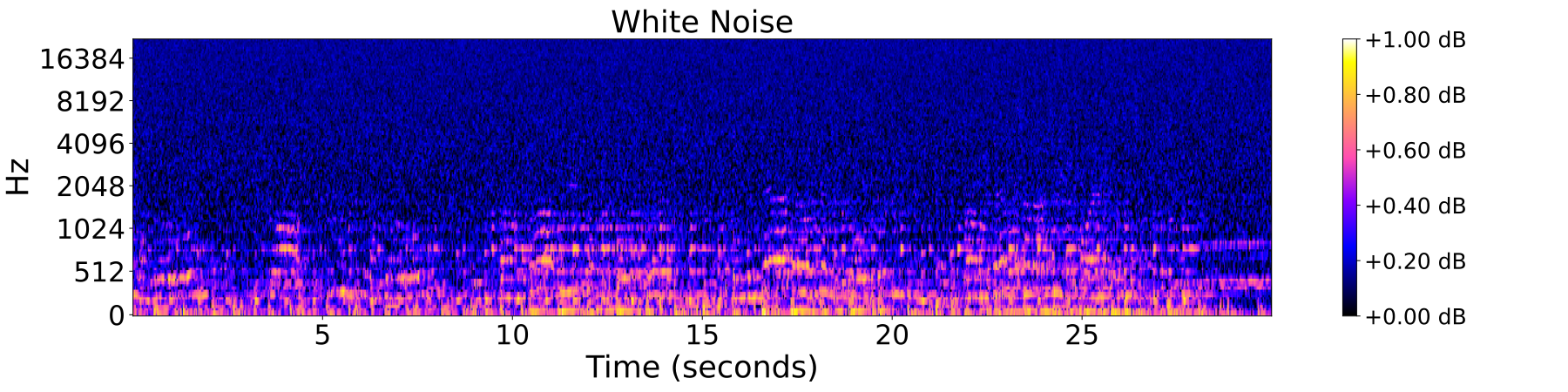} 
\includegraphics[width=0.25\textwidth]{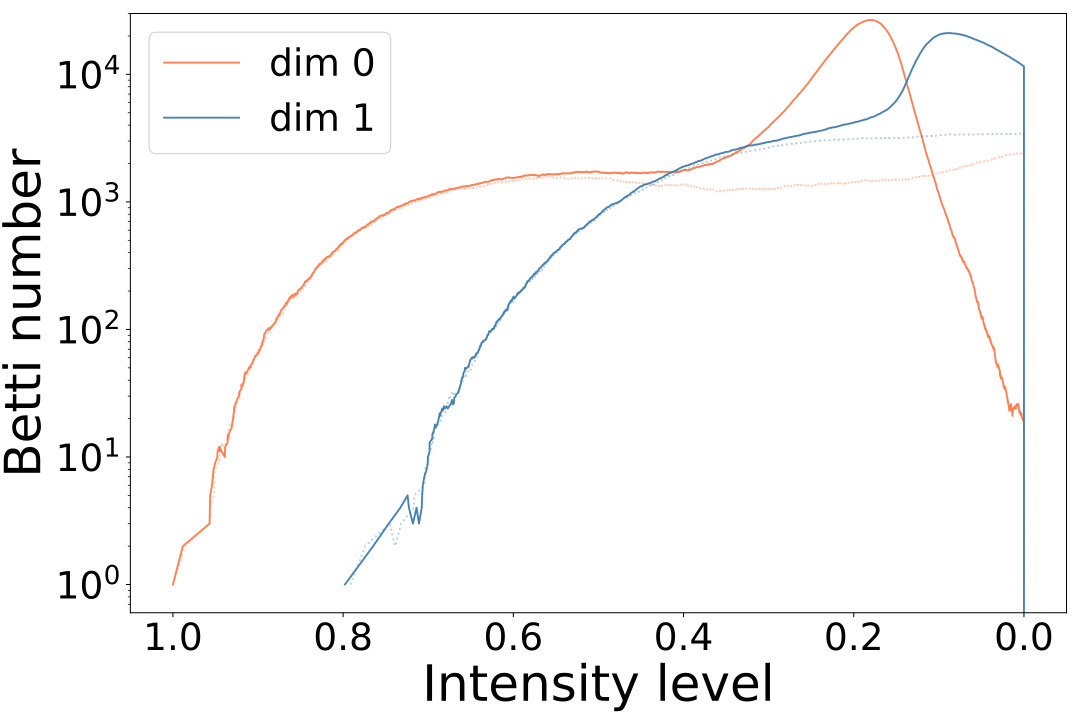}

\includegraphics[width=0.71\textwidth]{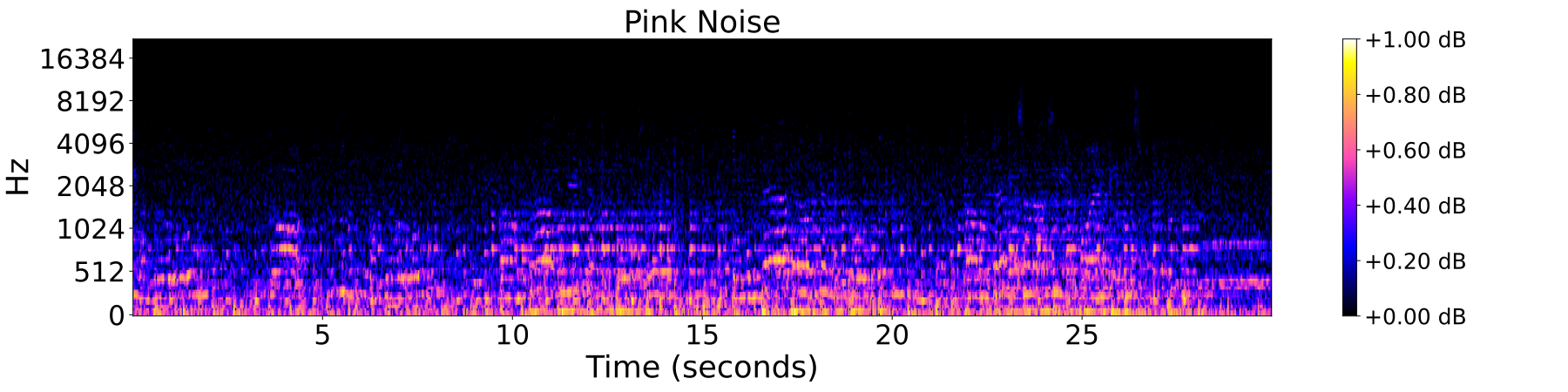} 
\includegraphics[width=0.25\textwidth]{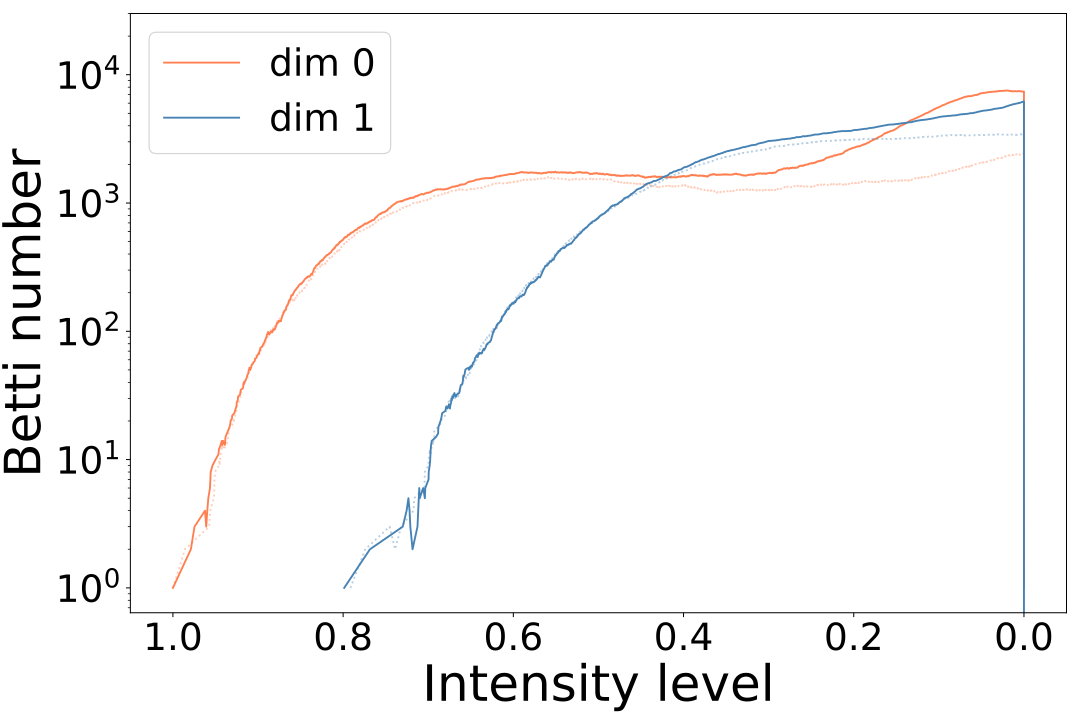}

\includegraphics[width=0.71\textwidth]{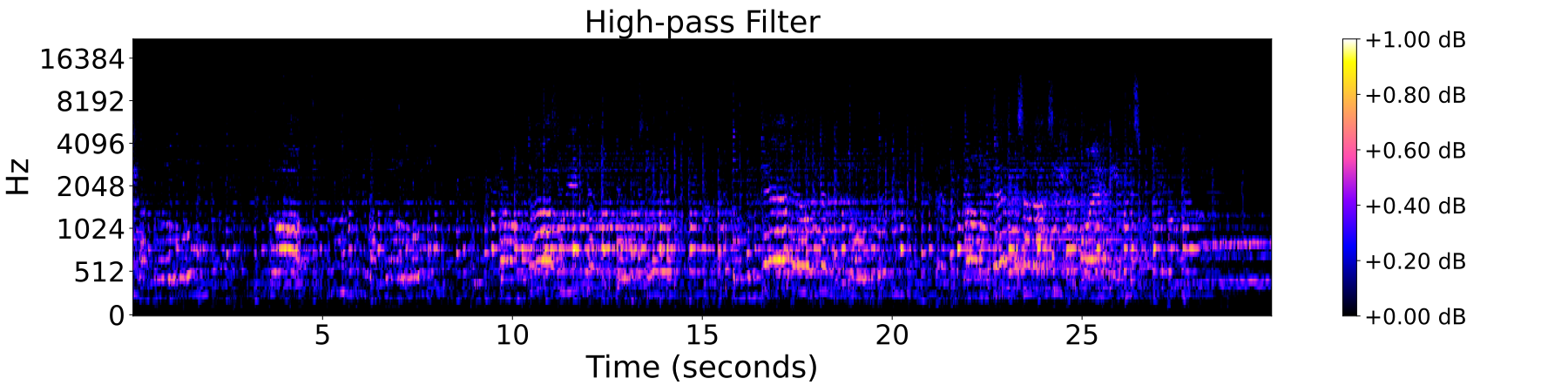} 
\includegraphics[width=0.25\textwidth]{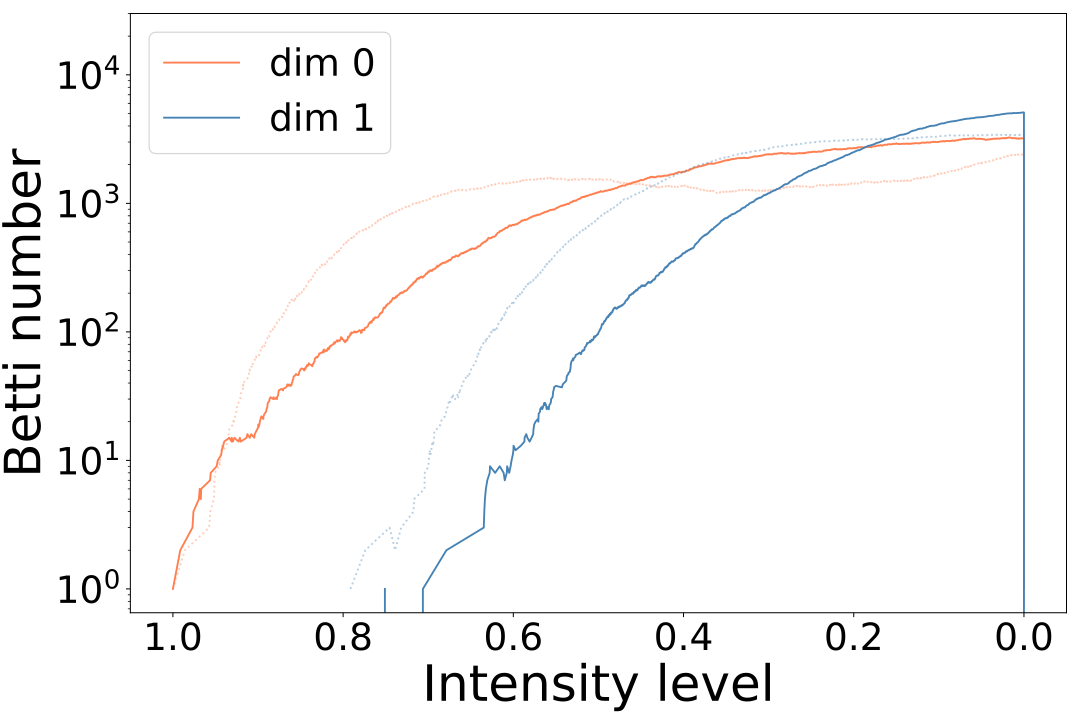}

\includegraphics[width=0.71\textwidth]{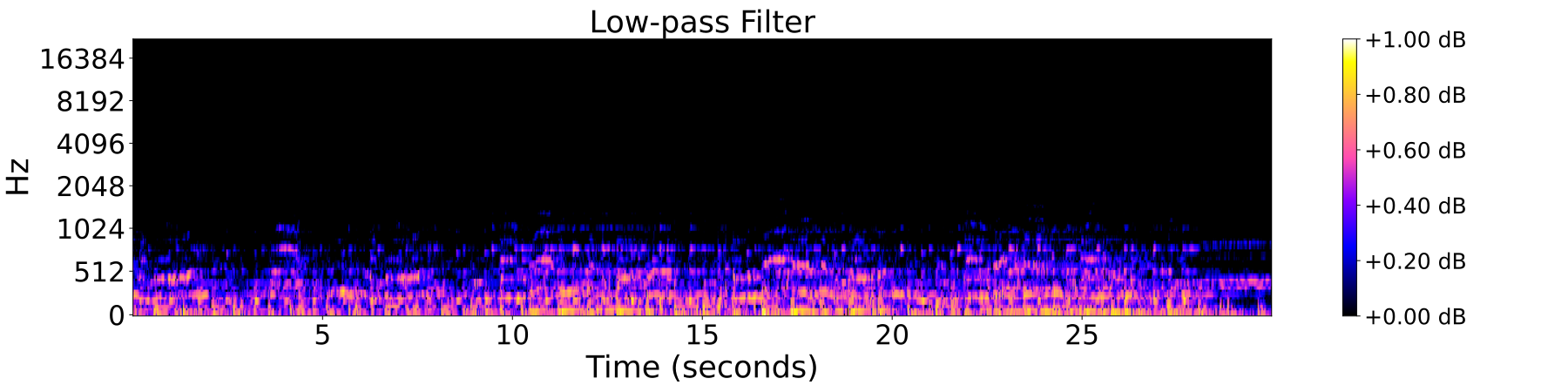} 
\includegraphics[width=0.25\textwidth]{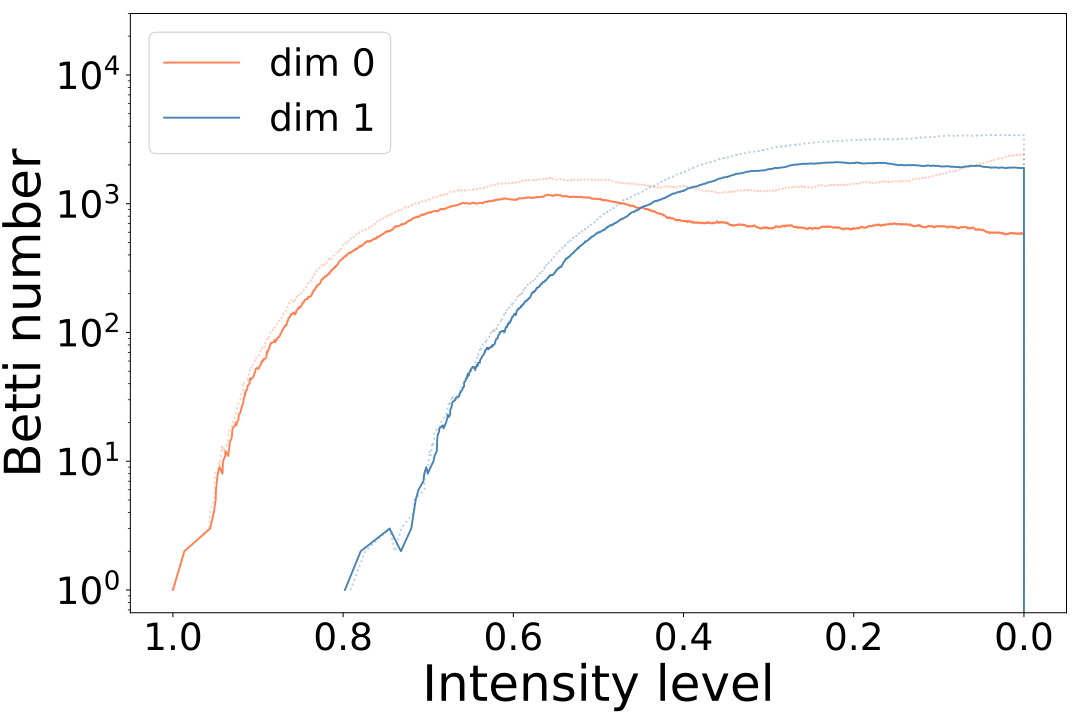}

\includegraphics[width=0.71\textwidth]{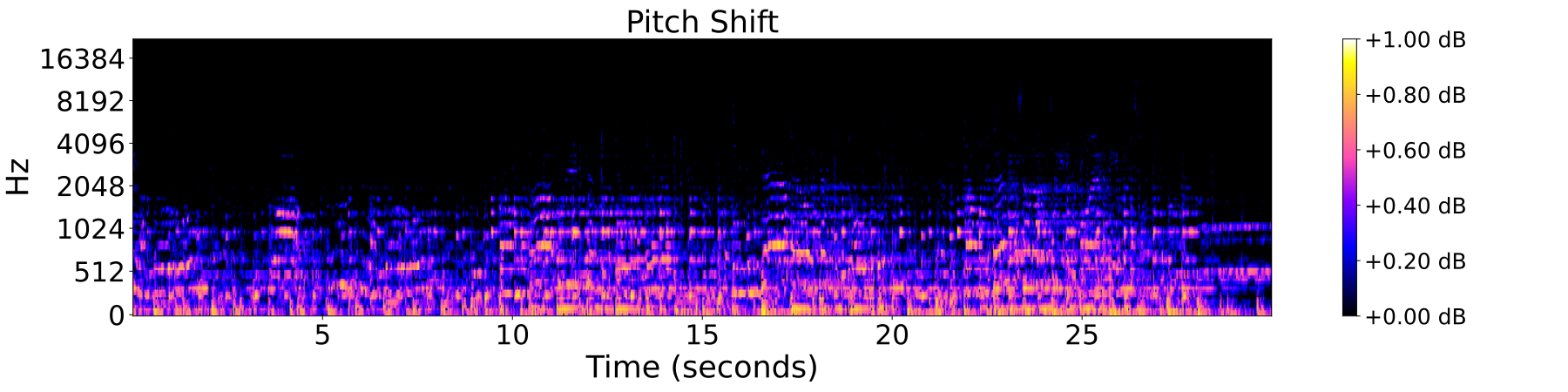} 
\includegraphics[width=0.25\textwidth]{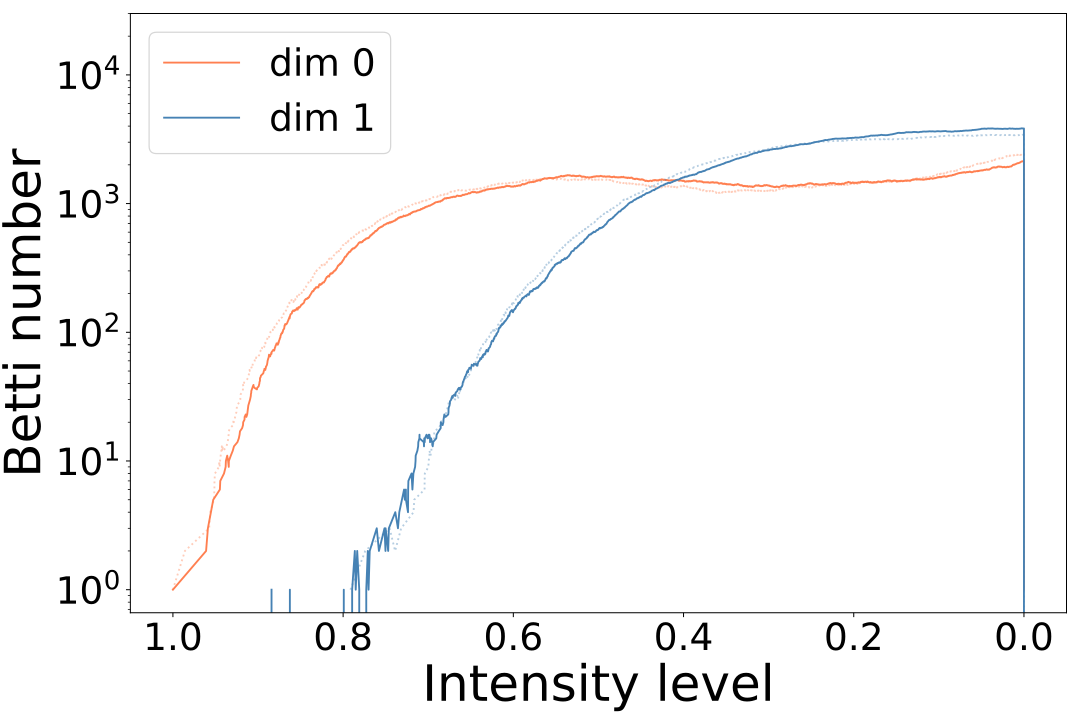}

\includegraphics[width=0.71\textwidth]{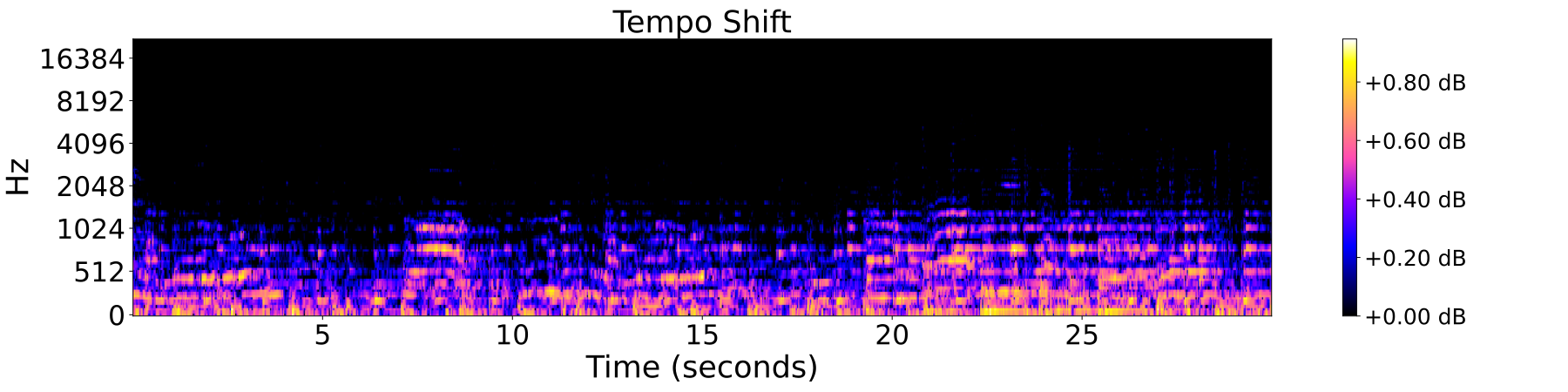} 
\includegraphics[width=0.25\textwidth]{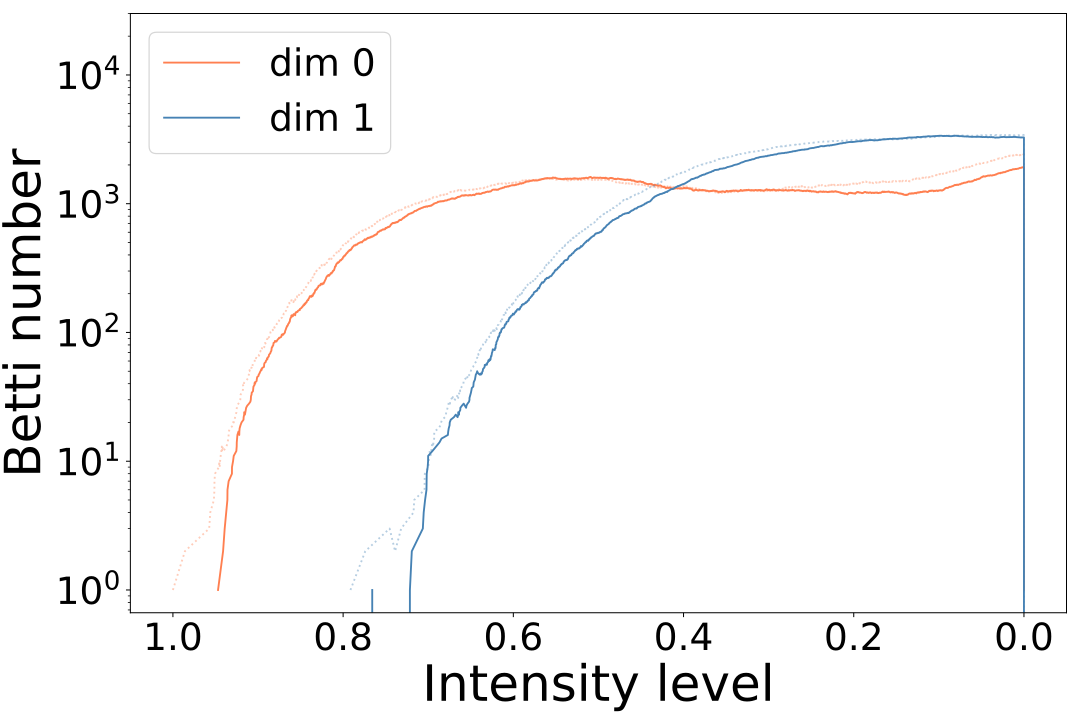}

\includegraphics[width=0.72\textwidth]{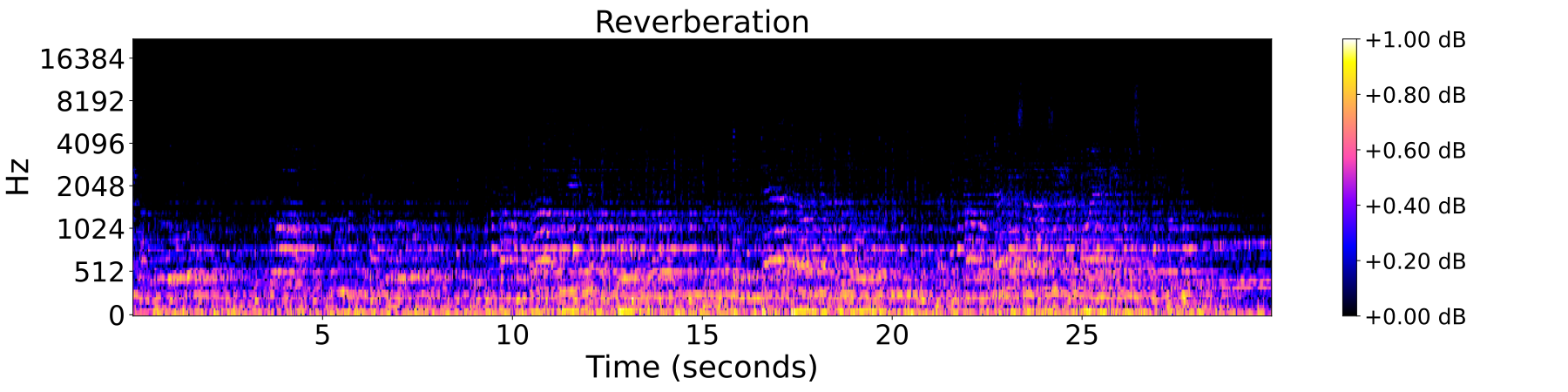} 
\includegraphics[width=0.25\textwidth]{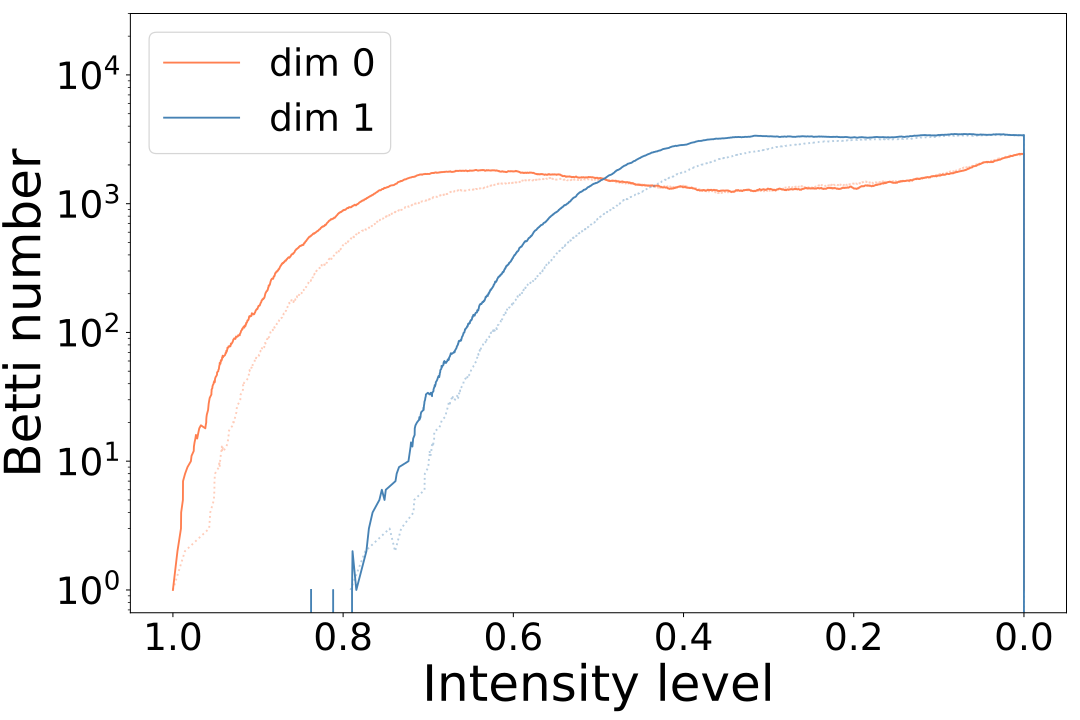}

\caption{\textbf{Betti curves of audio obfuscations.}  \textit{Left:} Mel-spectrograms of a 30-seconds fragment of the song  `The Morning' by \textit{Le Loup} \cite{le_loup_morning_2019} (c.f. \cref{fig:waveform_spectrogram}) and  \textit{obfuscated} versions. \textit{Right:} Betti curves of the associated persistence barcodes (in solid line) and the reference of the Betti curve of the original song (in light line).} \label{fig:obfuscations_topology}
\end{figure}

Audio obfuscations induce particular changes in the Betti curves associated to its spectral decompositions. 
Consider the example in \cref{fig:obfuscations_topology} of an original fragment of the song  `The Morning' by \textit{Le Loup} \cite{le_loup_morning_2019} (c.f. \cref{fig:waveform_spectrogram}) and its obfuscated versions. 
In the case of addition of noise (either white or pink), the noisy and  original versions of the track present a similar Betti curve for higher values of the intensity parameter (where the noise effect is less recognizable in the spectrogram), whereas there is noticeable a change in the Betti values for low intensities.  As result, the $L_1$-norm between the Betti curves for time-aligned windows remains small.
High and low-pass filters eliminate spectral information of high and low frequencies respectively, deriving in a decrease in the Betti curves in for the high and low interval frequencies respectively. In the example, we performed a high-pass and low-pass filter with cutoff frequency equal to 400 Hz. Pitch and tempo obfuscations distort the spectrogram in the frequency and time coordinates respectively. In the example, we performed a pitch shift of 8 semitones and a time stretch of ratio 0.5.
The topological information encoded in the Betti curves remains similar.

\section{Topological audio ID algorithm}
\label{sec:fingerprints}
In this section, we present
our method for topological audio identification based on persistent homology. 
We describe the key components of our approach, including the computation of persistent homology summaries on local spectral features (topological fingerprints) and the development of an audio identification algorithm for pairwise comparisons (matching algorithm).

\subsection{Fingerprinting audio tracks}
\label{sec:fingerprints_spectrogram}
Let $\mathcal S$ denote the mel-spectrogram of an audio track $s:[0,T]\to \mathbb{R}$. To obtain a local fingerprint of the audio signal, we subdivide $\mathcal S$ into a set of $\tau$-overlapping windows $W_0, W_1, \dots, W_k$ of $\omega$ seconds duration using a window size $\omega$ and an overlap constant $0<\tau<1$. Here, $W_i$ is the spectrogram of the audio track restricted to the time interval $\mathcal{I}_i = [i(1-\tau)\omega, i(1-\tau)\omega + \omega]$ for $0\leq i \leq k$, with $k\sim\frac{T-\omega}{\omega(1-\tau)}$. Let $t_i$ be the mid-point of $\mathcal{I}_i$. We normalize the range of intensity values of every window as
\[\frac{W_i-\min(W_i)}{\max(W_i)-\min(W_i)}.\]
Now, the procedure follows as in \cref{sec:topology_spectrograms}. For each (normalized) window $W_i$, we compute the persistent homology of the associated upper-filtered cubical complex for dimensions 0 and 1. Finally, we encode the persistent barcodes as a family of Betti curves $\{\beta_{i,0}\}_{i=0}^{k}$ and $\{\beta_{i,1}\}_{i=0}^{k}$ for dimensions 0 and 1, respectively.
The \textit{topological fingerprint} of the audio track $s$ with the resolution given by the parameters $\omega$ and $\tau$ is defined as the set of triples ${(t_i, \beta_{i,0}, \beta_{i,1})}_{i=0}^{k}$.

\subsection{Comparing audio tracks}
\label{sec:compare_fingerprints}

Let $s$ and $s'$ be two audio tracks. Given a window size $\omega$ and an overlap constant $\tau$, we subdivide the associated spectrograms $\mathcal{S}$ and $\mathcal{S}'$ into collections of $\tau$-overlapping windows $\{W_i\}_{i=0}^{k_s}$ and $\{W_j'\}_{j=0}^{k_s'}$ of size $\omega$ and compute their respective topological fingerprints$\{(t_i, \beta_{i,0},\, \beta_{i,1})\}_{i=0}^{k_s}$ and $\{(t_j', \beta'_{j,0},\, \beta'_{j,1})\}_{j=0}^{k_s'}$. 

For every homological dimension $d=0,1$, the \textit{$d$-Betti distance matrix} $M_d\in \mathbb{R}^{(k_s+1)\times (k_s'+1)}$ between $s$ and $s'$ is defined as
\begin{equation*}
(M_d)_{i,j} = \Vert \beta_{i,d} - \beta'_{j,d} \Vert_{L^1},
\end{equation*}
with $\Vert \cdot \Vert_{L^1}$ denoting the $L^1$-norm. Next, we summarize the distance between every pair of windows $W_i$ and $W_j'$ using a weighted combination of the $d$-Betti distances, with weights determined by a parameter $\lambda \in [0,1]$, giving us
\begin{equation} \label{eq:cost_matrix}
C_{i,j} = \lambda (M_0)_{i,j} + (1-\lambda) (M_1)_{i,j}.
\end{equation}

We compare the spectrograms $\mathcal{S}$ and $\mathcal{S}'$ via a minimum-cost matching in $C\in \mathbb{R}^{(k_s+1)\times (k_s'+1)}$ using the Hungarian algorithm \cite{kuhn_hungarian_1955}\footnote{This problem is also known as the \textit{assignment problem} in combinatorial optimization and can be solved in both the balanced and unbalanced cases in polynomial time.}. The output of this procedure is a binary matrix $X \in \mathbb{R}^{(k_s+1)\times (k_s'+1)}$ with $X_{i,j} = 1$ if the window $W_i$ centred at $t_i$ is paired with the window $W_j'$ centred at $t_j'$. Note that the matching $X$ may not respect the temporal relation between the set of matched windows.

\begin{figure}[htb!] 
	\hspace{-35pt}	\includegraphics[width=0.42\textwidth]{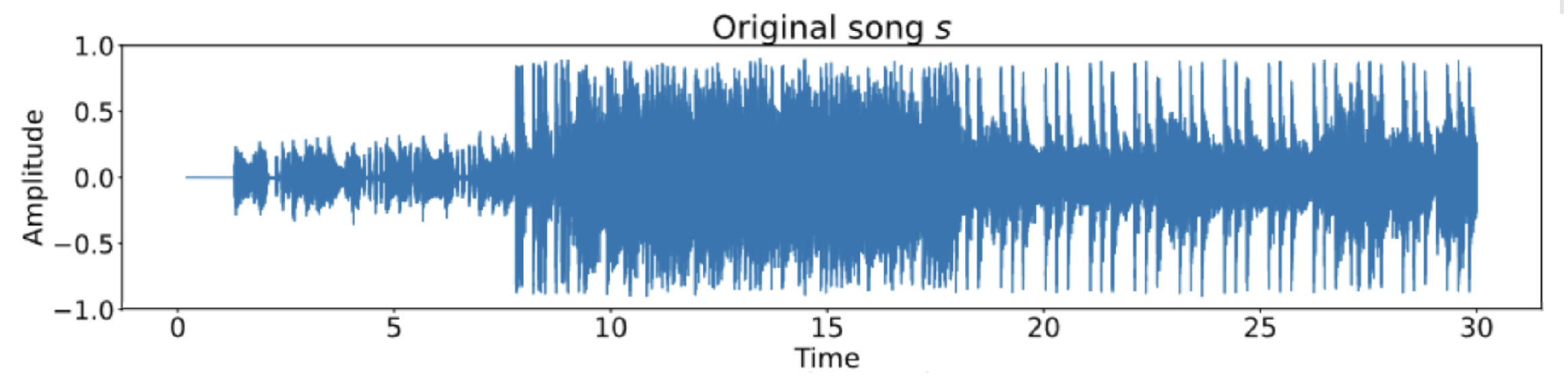} \hspace{25pt}	\includegraphics[width=0.42\textwidth]{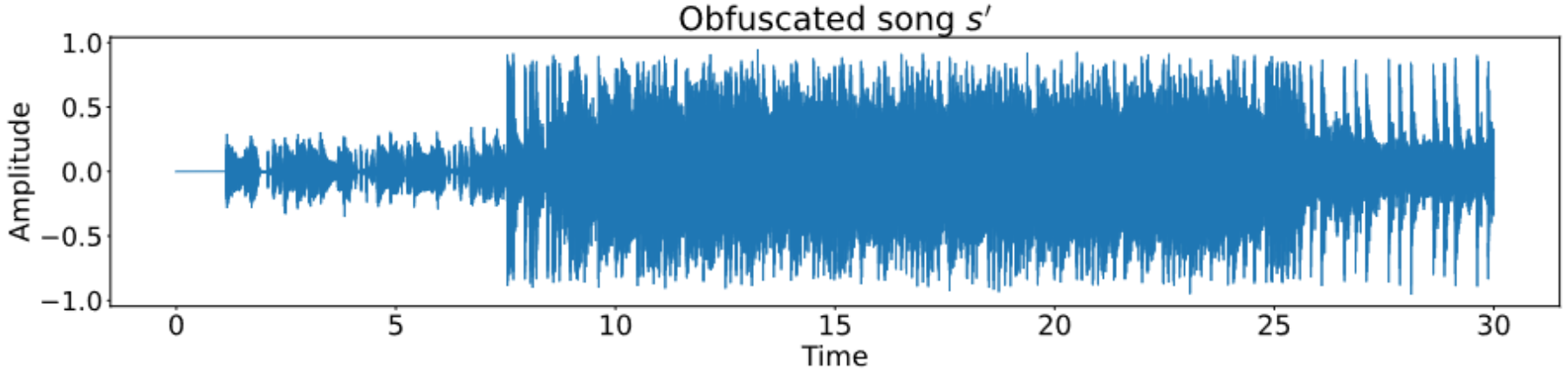}

	\includegraphics[width=0.49\textwidth]{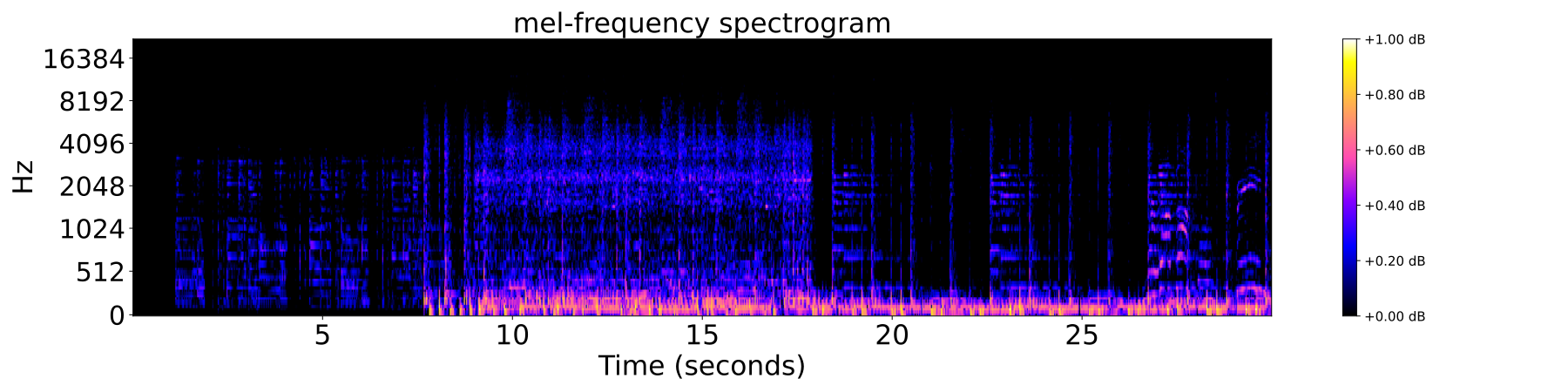}	\includegraphics[width=0.49\textwidth]{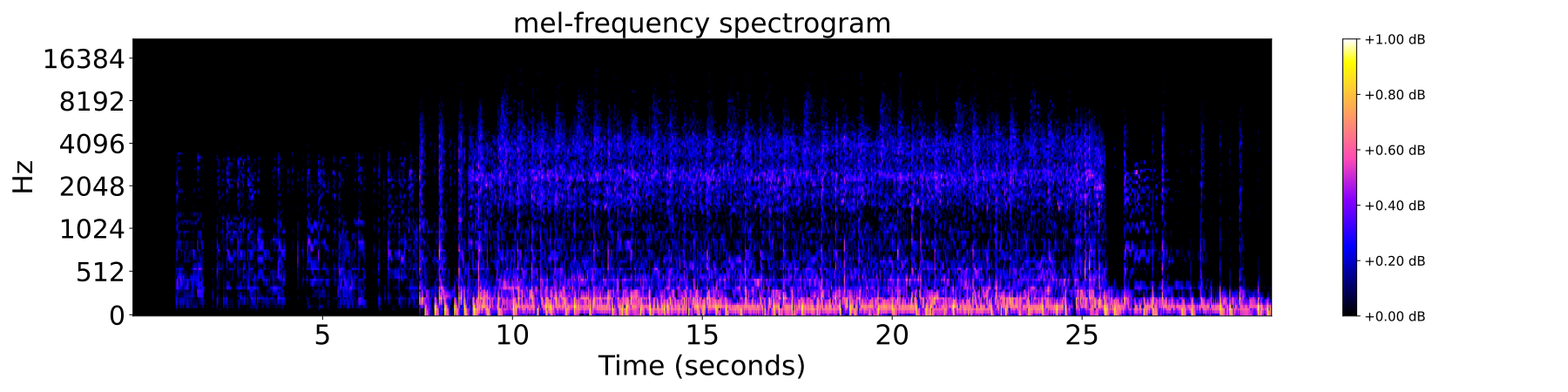}

	\includegraphics[width=0.49\textwidth]{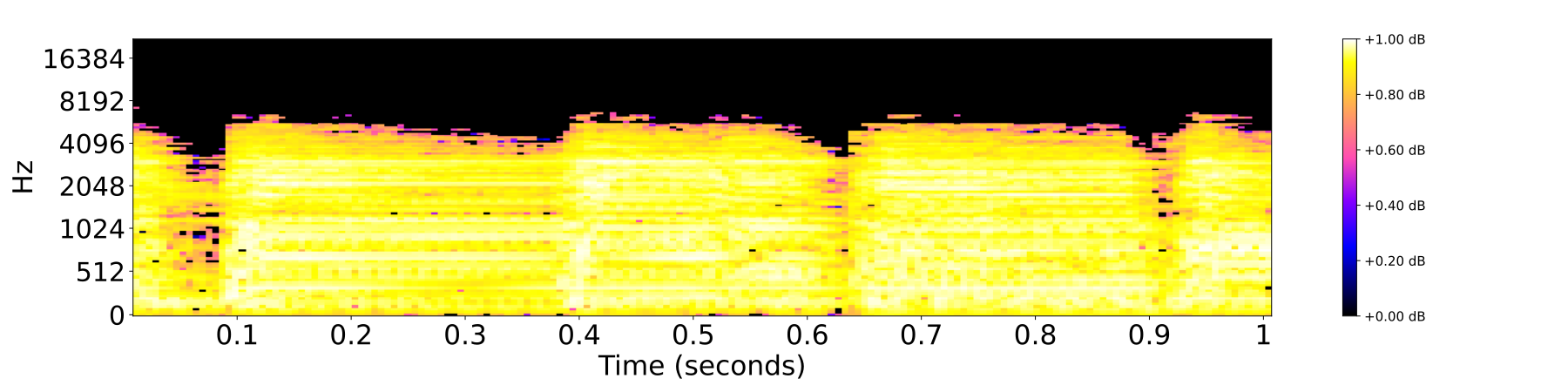}	\includegraphics[width=0.49\textwidth]{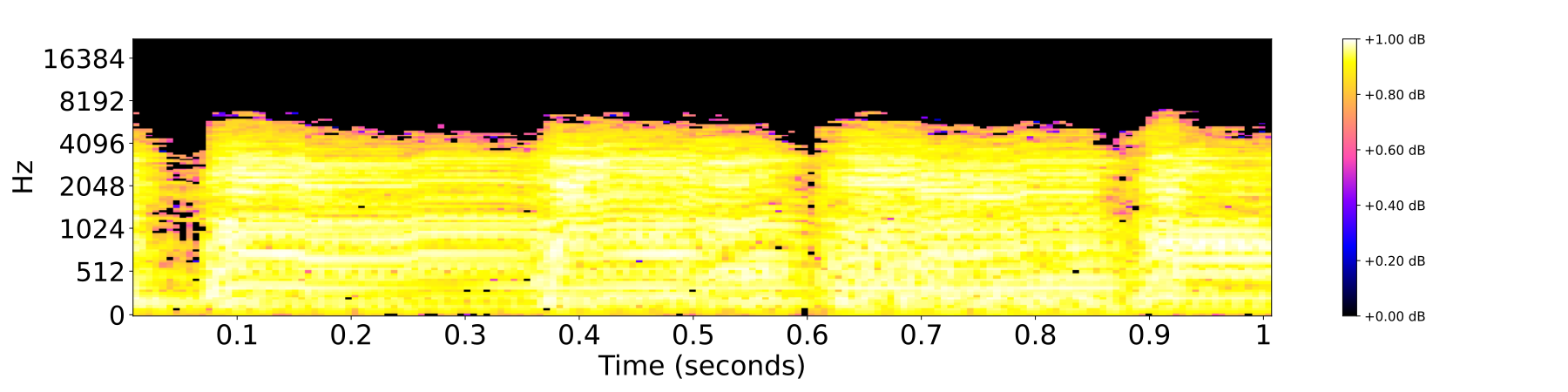}

 	\hspace{-35pt}\includegraphics[width=0.4\textwidth]{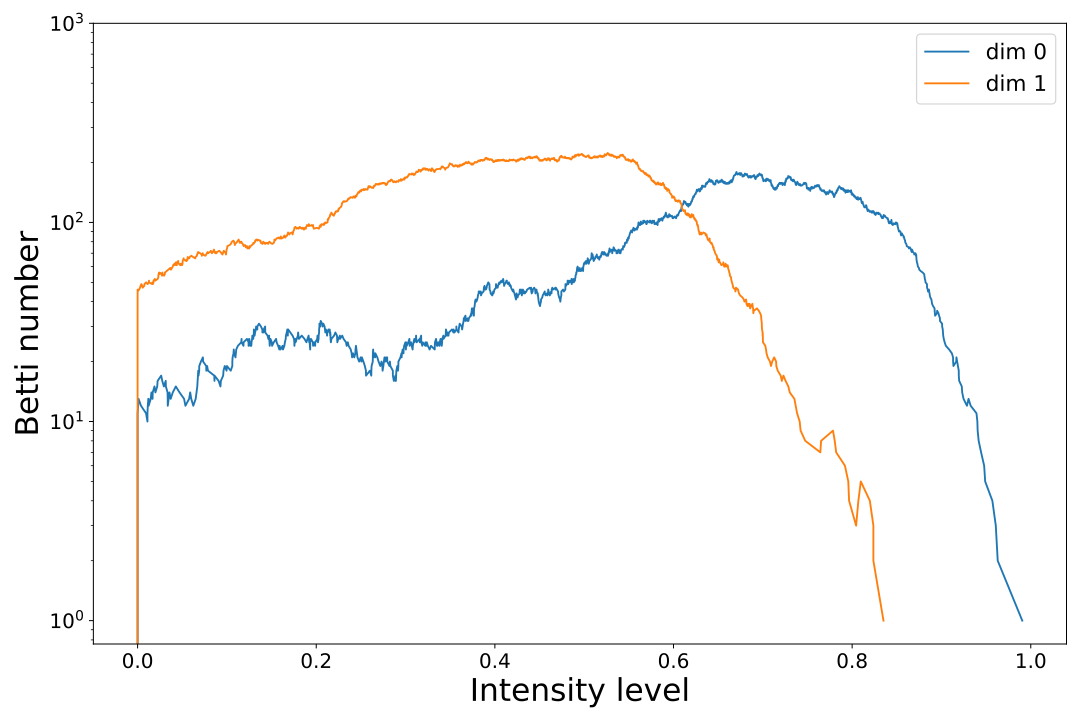}	\hspace{35pt}	\includegraphics[width=0.4\textwidth]{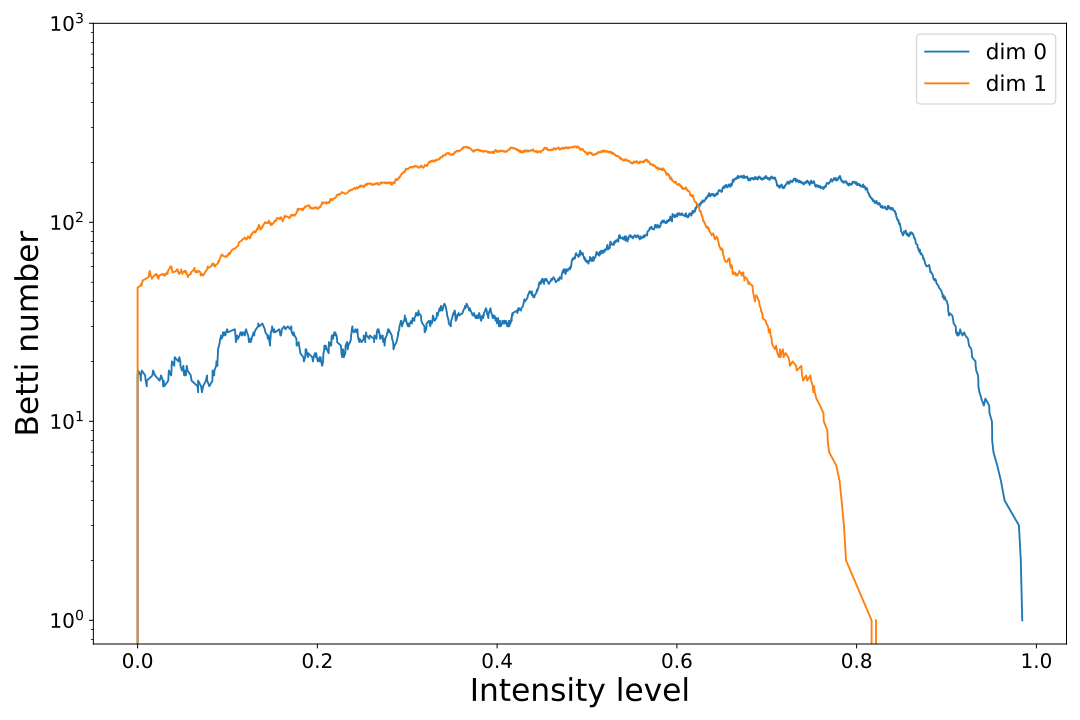}

    \hspace{-43pt}\includegraphics[width=0.87\textwidth ]{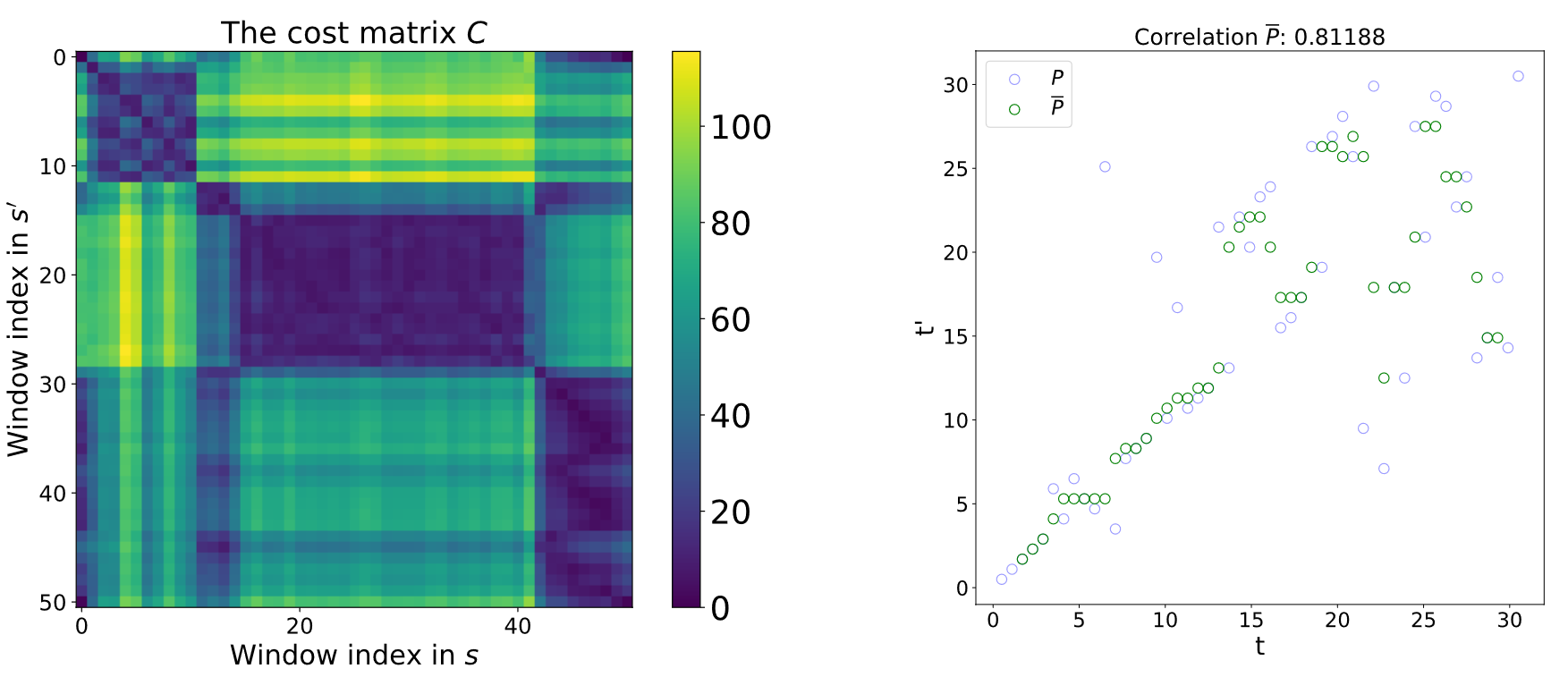}

 \caption{\textbf{The topological audio ID algorithm.} \textit{Top.} The waveform and the mel-spectrograms of 30 seconds of the audio track `Smells like teen spirit' by \textit{Nirvana} \cite{smells1991} (left) and its obfuscated version by Ben Grosser \cite{grosser_music_obfuscator} (right). \textit{Middle:} 1-second window extracts (normalized) and its topological fingerprints given by the Betti curves of dimensions 0 and 1. \textit{Bottom:} The cost matrix $C$ for $\lambda = 0.3$ (left). The graph of the points $P\subseteq \mathbb{R}^2$ of pairs of times $(t_i,t_{i_j})$ of matched pairs of windows $(W_i, W_{i_j})$ with respect to a minimum-cost matching in blue, and the set $\overline{P}$ of the moving average of the points in  $P$ in green (right).}
 \label{fig:smells_like_teen_spirit}
\end{figure}

To measure the similarity between two tracks $s$ and $s'$, we need to quantify the degree of temporal-order preservation in the matching $X$. 
 Let $P = \{(t_1, t_{j_1}),  (t_2, t_{j_2}), \\ \dots, (t_n, t_{j_n})\mid X_{i,j_i} = 1\}$ be the set of mid-points of matched windows, with $t_1<t_2<\dots<t_n$.
 The ideal situation of perfectly aligned matched windows corresponds to the case $t_{j_1}<t_{j_2}<\dots,t_{j_n}$ (that is, the function $t_i\mapsto t_{j_i}$ is an increasing function). 
We assess the level of increasing relationship between the points in $P$ as 
\begin{equation*}
\rho_P = \mathrm{Pearson}(\{t_i\}_{i=1}^n, \{\bar t_{i_j}\}_{i=1}^n).
\end{equation*}
Here, we replaced $t_{i_j}$ by its \textit{neighborhood median} $$\bar t_{i_j} = \median \{t_{j_{i-k}}, \dots, t_{i_{j-1}}, t_{i_j}, t_{i_{j+1}},\\ \dots, t_{j_{i+k}}\},$$ for some parameter $k>0$ to increase robustness.
Notice that $\rho_{P}$ is close to 1 when there is a time-preserving dependency between the points in (the \textit{smoothed} version of) $P$. Hence, we define the error as
\begin{equation}
E(s,s') = 1-\rho_P.
\label{eq:compare_fingerprints_error}
\end{equation}
Consider the example in \cref{fig:smells_like_teen_spirit} of the  song  `Smells like teen spirit' by \textit{Nirvana} \cite{smells1991} and its obfuscated version developed by Ben Grosser \cite{grosser_music_obfuscator} (which was not recognized by the Shazam application in 2016). 
We compare 30 seconds extracts from the original song $s$ and the obfuscated version $s'$. We extract the associated Betti curves from 1-second overlapping windows of the associated spectrograms (with an overlap constant of 0.4 seconds). For  $\lambda=0.3$,
the optimal time-matching between the windows has an error of $1-\rho_P = 0.188122$, which suggests that the pair of audio tracks are likely to correspond the same audio content. For more rigorous experiments on this example see \cref{sec:experiments}.

\section{Experimental results}\label{sec:experiments}
In this section, we evaluate the effectiveness of the fingerprinting and comparison methods described in \cref{sec:fingerprints} for identifying audio tracks with the same audio content. Our goal is to determine whether a pair of tracks $s$ and $s'$ are duplicates, with one of the tracks being modified through obfuscations to yield the other. We present experimental results showcasing the performance of our algorithm on a variety of obfuscation types. While our method shows similar performance to existing approaches under \textit{rigid} obfuscations, such as the addition of noise or reverb, it significantly outperforms them in cases where the signal is subject to \textit{topological} deformations, such as time stretching or pitch shifting.

\subsection{Datasets} We randomly selected  around $135,000$ 
tracks from the Million Song Dataset \cite{bertin_million_2011} and  downloaded 30 second preview snippets using the {\fontfamily{lmss}\selectfont Spotify Web-API} \cite{spotifywebapi}. All the audio files $s$ are sampled at frequency $f_s = 44.1kHz$.
Every audio track has been then manipulated according to seven types of obfuscations  of different levels of magnitude.
The degree determines to what extent the track is distorted. For example, a tempo shift with a factor of $1.05$ indicates that the track has been sped up by a factor of $0.05$, without changing its pitch.
Obfuscations, summarized in \cref{tbl:obfuscations}, have been generated using the Python  wrapper  {\fontfamily{lmss}\selectfont  pysox} \cite{bittner_pysox:_2016} of the audio editing software {\fontfamily{lmss}\selectfont  SOX}. 

\begin{table}[htb!]
	\centering
	\begin{tabular}{|l|l|}
	    \hline
		\textbf{Type} & \textbf{Degree}\\
		\hline
		Low-pass filter& 200, 400, 800, 1600, 2000 \\
		High-pass filter&  50, 100, 200, 400, 800, 1200\\
		White noise & 0.05, 0.1, 0.2, 0.4\\
		Pink noise & 0.05, 0.1, 0.2, 0.4\\
		Reverb & 25, 50, 75, 100\\ 
		Pitch shift& -8, -4, -2, -1, 1, 2, 4, 8\\
		Tempo shift& 0.5, 0.8, 1.1, 1.2, 1.5, 2.0\\
		\hline
	\end{tabular}
\caption{\textbf{Audio obfuscations.} Types of transformations and degrees used to generate the dataset of obfuscated tracks. In low (resp. high) pass filter, the degree is the threshold frequency, so the higher the threshold, the smaller (resp. greater) the distortion. For white noise, pink noise and reverb,  the degree measures the amount of the effect added to the original signal (percentage in the case of reverberation). For pitch shift, the degree quantifies the number of semitones in any direction that are continuously distorted.  Finally, 
for tempo shift, the degree is the factor of the time stretching, being $1$ the identity.
}
\label{tbl:obfuscations}
\end{table}

If $s$ is an original audio track  and $s'$ is an obfuscated version of $s$, then $(s,s')$ constitutes a \textit{positive pair}. A \textit{negative pair} is any pair of (possibly obfuscated) different audio tracks. For every obfuscation type and degree, we sampled 1,000 original songs and obfuscated pairs from the dataset, generating a set of 36,000 positive pairs. We also randomly selected 36,000 pairs of (possibly obfuscated) but unrelated songs.
We equally split the generated dataset of 72,000 pairs of songs into training and test sets.

A second dataset was created using the challenging examples provided by Ben Grosser in his work on the \textit{Music Obfuscator} \cite{grosser_music_obfuscator}. The dataset consists of 8 tracks spanning different music genres, which were manipulated using Grosser's signal processing algorithm. The Music Obfuscator algorithm \cite{grosser_music_obfuscator} applies tempo and pitch shifting with varying degrees over time to introduce distortions into the tracks. Remarkably, these manipulated tracks were undetectable by content ID algorithms on platforms like YouTube and Soundcloud as of 2016. Even in 2023, the current version of the Shazam app fails to recognize many of these tracks, despite their ease of recognition by humans.
Although the dataset is small, these challenging examples serve as a rigorous test for evaluating the performance of audio identification systems under complex obfuscations.  

\subsection{Implementation} Our \textit{Topological Audio ID Algorithm} is implemented in Python and available at \cite{topological_audioID}. The spectral features from audio tracks are computed with the  library {\fontfamily{lmss}\selectfont  librosa}  \cite{mcfee_librosa:_2015}. We use a window size $N_w = 1024$ for the short-time Fourier transform (and a hop size $h = 256$ for subsampling in time).
We decompose the frequency spectrum in $M=128$ bins, with $f_1 = \frac{1}{N_w}$ and $f_M = \frac{f_s}{2}$ the Rayleigh and Nyquist frequencies respectively.
Given an audio signal $s$ of length $T$ seconds, the resulting spectrogram $\mathcal{S}$ has $Tf_s/h$ columns and $M=128$ rows.
%
%
%
For the fingerprinting algorithm, we extract windows $W_i$ of size $w=1$ second, with overlap rate $\tau = 0.4$. 
Every sub-spectrogram $W_i$ is an image of size $172\times 128$. 
Cubical filtrations and persistent homology are computed with the library {\fontfamily{lmss}\selectfont{Gudhi}} \cite{gudhi:urm} for TDA. 
The minimum cost time-matchings are obtained with the method {\fontfamily{lmss}\selectfont{optimize.linear\_sum\_assignment}} from {\fontfamily{lmss}\selectfont{scipy}} \cite{scipy}.

To compare the performance of our algorithm with the popular algorithm \textit{Shazam}, we employ an open-source software (OSS) implementation available at \cite{shazam_open_source}, as well as the current available version of the application \cite[Shazam Version 15.35.0 (5369:f)] {shazam_2023}.





\subsection{Results}
\subsubsection{Million song  dataset}\label{sec:spotify_results} 
\begin{figure}[htb!]
	\centering
	\includegraphics[width=0.8\textwidth]{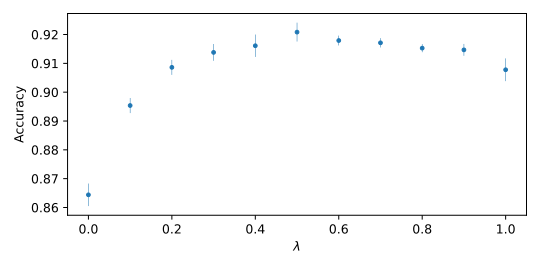}
	\caption{\textbf{The choice of the parameter $\lambda$.} The distribution of the accuracy of our method for a range of values of $\lambda\in[0.1]$. The highest accuracy values is reached for $\lambda=0.5$.}
	\label{fig:cross_val_lambda}
\end{figure}

 For every labeled pair of songs $(s,s')$, the \textit{Topological Audio ID} algorithm outputs an error score $E_{(s,s')}$ \eqref{eq:compare_fingerprints_error}. We set the value of the smoothing constant $k=2$.
Notice that the topological error \eqref{eq:compare_fingerprints_error} strongly depends on a parameter $\lambda$ that rules the linear combination \eqref{eq:cost_matrix} of the cost matrices associated to Betti curves of dimension 0 and 1.
To determine the optimal value of the parameter $\lambda$, we perform a cross-validation analysis with 4 folds during the training phase using logistic regression. We then measure the average accuracy of our algorithm over the folds (refer to \cref{fig:cross_val_lambda}). The best value of $\lambda$ is found to be 0.5, assigning equal importance to 0 and 1 dimensional homological features.

\begin{figure}[htb!]
\includegraphics[width = \textwidth]{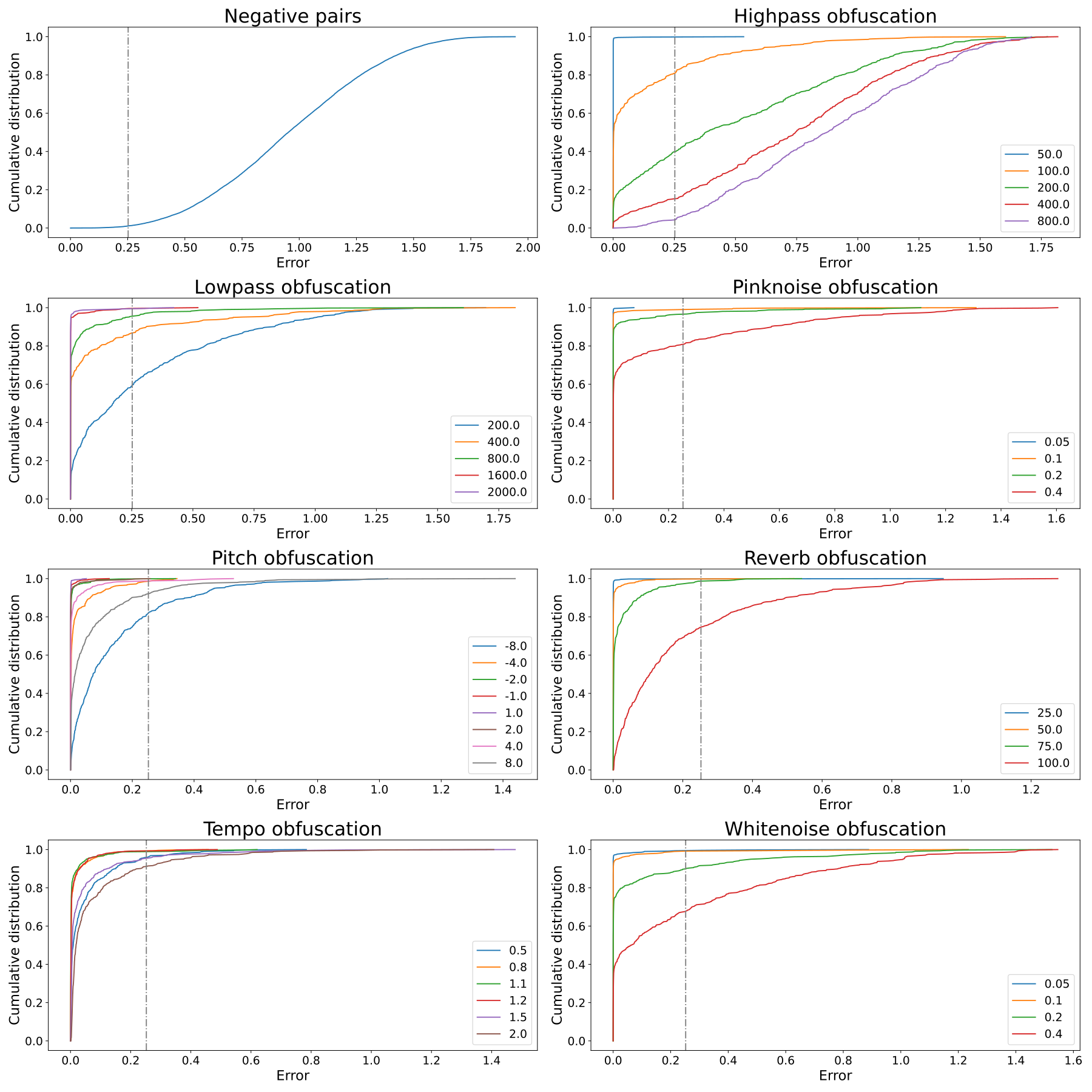}
\caption{\textbf{Cumulative distribution.} For negative pairs and positive pairs (grouped by obfuscation type), the cumulative distribution of pairs whose error is less than or equal to a threshold in $[0,2]$. 
In dashed line, the optimal threshold $\kappa_{\mathrm{Top}}$.}\label{fig:cummulative_TDA}
\end{figure}

Next, we perform the computation of the error score \eqref{eq:compare_fingerprints_error} over the whole dataset using the optimal value of $\lambda$. A threshold $\kappa_{\mathrm{Top}}$ for the classification of positive and negative pairs is set in order to achieve a \textit{false positive rate} of 0.01 on the training set. This means that if we classify a pair $(s, s')$ as positive whenever $E(s,s')<\kappa_{\mathrm{Top}}$, then we expect to incorrectly classify at most $1\%$ of positive pairs.
A summary of the performance scores on the test set is presented in \cref{tbl:scores}.  A visual representation of the cumulative distribution of pairs across various threshold values for the different obfuscation types is presented in \cref{fig:cummulative_TDA}.

\begin{figure}[htb!]
	\centering
    \includegraphics[width = 0.65\textwidth]{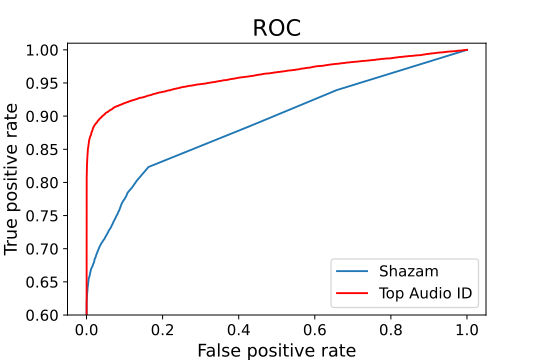}
    
	\caption{\textbf{Comparison of ROC curves.}  Detail of the Receiver Operating Characteristic (ROC) curves for our Topological Audio ID algorithm (red) and Shazam (OSS) (blue). 
    The area under the ROC curve is  $0.929854$ for our method and  $0.828585$ for Shazam.}
	\label{fig:ROC_curves}
\end{figure}

A similar experiment for \textit{Shazam}  is reproduced using its open-source software (OSS) implementation \cite{shazam_open_source}. For the \textit{$1$ vs $1$} comparison purposes, we returned the score that counts the maximum number of time aligned constellation hashes (see \cite[Sect. 2.3]{wang_industrial-strength_2003}).
Again, a threshold $\kappa_{\mathrm{Sh}}$ for the classification task is set to achieve a false positive rate of 0.01 on the training set.
The performance scores for Shazam (OSS) can be read from  \cref{tbl:scores}, while
the comparison of the ROC curves for both methods is depicted in \cref{fig:ROC_curves}. We observe an overall superior performance of our method over the benchmark, with high values of both \textit{precision} (accuracy among pairs classified as positive pairs) and \textit{recall} (rate of positive pairs that were correctly identified). 

\begin{table}[htb!]
	\centering
	\begin{tabular}{|l|r|r|}
	    \hline
		\textbf{Score} & 
        \textbf{Shazam (OSS)} & \textbf{Top. Audio ID}\\
		\hline
		Accuracy&   0.8290 & {\bf 0.9300}\\
		Precision & 0.9829 & {\bf 0.9874}\\
		Recall &    0.6686 & {\bf 0.8708}\\
		AUC &       0.8286 & {\bf 0.9299}\\
		\hline
	\end{tabular}
\caption{\textbf{Classification results for the Million Song dataset.} In bold the highest scores. If we denote $\TP$ (resp. $\TN$) the rate of positive (resp. negative) pairs correctly identified and $\FP$ (resp. $\FN$) the rate of incorrectly classified pairs, then \textit{accuracy} is defined as $\frac{\TP+\TN}{\TOT}$, \textit{precision} is $\frac{\TP}{\TP+\FP}$ and \textit{recall} is $\frac{\TP}{\TP+\FN}$. The \textit{area under the curve} (AUC) can be visualized in \cref{fig:ROC_curves}.}
\label{tbl:scores}
\end{table}

A comprehensive analysis of the performance of our algorithm compared to the benchmark algorithm for different obfuscation types is presented in \cref{tbl:results_by_obfuscation}. 

\begin{table}[htb!]
	\centering
	\begin{tabular}{|l|r|r|r|}
		\hline
		\textbf{Obfuscation type} & \textbf{Degree}  &  \textbf{Shazam (OSS)}
         & \textbf{Top. Audio ID}\\
		\hline
		High-pass filter
        & ~50 & {\bf 1.0000}   & 0.9980 \\
		& 100 & {\bf 1.0000}   & 0.8095 \\
		& 200 & {\bf 1.0000}   & \textcolor{red}{0.3996} \\
		& 400 & {\bf 1.0000}   & \textcolor{red}{0.1543} \\
        & 800 & {\bf 1.0000}   & \textcolor{red}{0.0429} \\
        \hline
        Low-pass filter  
        & ~200  & {\bf 0.9979} & \textcolor{red}{0.5968} \\
		& ~400  & {\bf 1.0000} & 0.8669 \\
		& ~800  & {\bf 1.0000} & 0.9552 \\
		& 1600  & {\bf 1.0000} & 0.9958 \\
  		& 2000  & {\bf 1.0000} & 0.9939 \\
        \hline
		Pink noise 
        & 0.05 & 0.9979       &  {\bf 1.0000} \\
		& 0.10 & {\bf 0.9978} &  0.9894 \\
		& 0.20 & {\bf 0.9902} &  0.9649  \\
		& 0.40 & {\bf 0.9599} &  0.8080  \\
        \hline
        White noise 
        & 0.05 & 0.9940       & {\bf 0.9940} \\
		& 0.10 & {\bf 0.9920} & {\bf 0.9920}     \\
		& 0.20 & {\bf 0.9246} & 0.8975     \\
		& 0.40 & {\bf 0.7962} & \textcolor{red}{0.6770}\\
        \hline
		Reverb 
        & ~25   & {\bf 0.9980}  &  {\bf 0.9980}\\
		& ~50   & {\bf 0.9961}  &  0.9881        \\
  		& ~75   & {\bf 0.9980}  &  0.9866        \\
		& 100   & {\bf 0.9961}  &  0.7457\\
        \hline
        Pitch shift
        & -8   & \textcolor{red}{0.0000}  & {\bf 0.8195}  \\
		& -4   & \textcolor{red}{0.0000}   & {\bf 0.9841} \\
		& -2   & \textcolor{red}{0.0000}   & {\bf 0.9980} \\
		& -1   & \textcolor{red}{0.0000}   & {\bf 1.0000}  \\
		& ~1   & \textcolor{red}{0.0000}   & {\bf 1.0000}  \\
		& ~2   & \textcolor{red}{0.0000}   & {\bf 0.9980}  \\
  		& ~4   & \textcolor{red}{0.0000}   & {\bf 0.9876}  \\
      	& ~8   & \textcolor{red}{0.0000}   & {\bf 0.9198}  \\
        \hline
		Tempo shift
        & 0.5  & \textcolor{red}{0.0675}  &  {\bf 0.9583} \\
		& 0.8  & \textcolor{red}{0.4333}  &  {\bf 0.9922} \\
		& 1.1  &                  0.9052  &  {\bf 0.9880} \\
		& 1.2  & \textcolor{red}{0.5210}  &  {\bf 0.9920} \\
		& 1.5  & \textcolor{red}{0.2653}  &  {\bf 0.9545} \\
		& 2.0  & \textcolor{red}{0.2141}  &  {\bf 0.9101} \\
		
		\hline
	\end{tabular}

	\caption{\textbf{Accuracy grouped by obfuscation type for the Million Song Dataset.} Comparison of the accuracy score for the proposed method against the benchmark, for the different obfuscation types and degrees. The highest score for every obfuscation type and degree is in bold, while values lower than 0.7 are marked in red. 
 }
	\label{tbl:results_by_obfuscation}
\end{table}

Our method demonstrates comparable results to Shazam (OSS) for obfuscations that introduce perturbations to the original signal, such as the addition of white noise, pink noise, and reverberation. However, for extreme levels of perturbation (e.g., a noise degree of 0.4 and a reverb level of 100), our performance noticeably declines due to the significant alteration in the topological fingerprints.

Notably, our approach excels in handling topological obfuscations. Unlike Shazam (OSS) \cite{shazam_open_source}, which struggles with tempo shifts and fails to recognize pitch-shifted songs, we achieve accuracy scores above 0.9 in almost all cases. It's worth mentioning that our accuracy scores are slightly lower for extremely high obfuscation degrees (specifically, 8 and -8 for pitch shift and 2 for tempo shift). These distortion levels also make it challenging for humans to recognize the obfuscated songs.

On the other hand, our method faces challenges with frequency filters. We achieve accuracy scores above 0.8 for low to moderate thresholds in high-pass and low-pass filters. However, for more extreme values, our detection accuracy diminishes. This issue is more pronounced with the high-pass filter, as it retains only the information from the higher regions of the spectrograms. Since most of the spectral power (and hence, geometric information) of songs is concentrated in low-frequency values, this type of obfuscation introduces significant differences in the Betti curves. 

These differences in the results are explained by the inherent characteristics of the fingerprinting and comparison algorithms of each method. While Shazam searches for time-aligned local peaks in the spectrogram, our algorithm looks for similar topological information in the sublevel sets of the associated surface. As a result, Shazam performs well even with minimal literal coincidences in the spectrograms (as observed in the case of high-pass filters), whereas our algorithm excels in cases where distortions displace local maxima but preserve the global shape. This distinction is particularly evident in Grosser's dataset \cite{grosser_music_obfuscator} (see \cref{grosser_results}).

\subsubsection{Grosser's dataset} \label{grosser_results}
We extract a 60-second fragment from each original song in Grosser's dataset (from 0 to 60 seconds), and a 30-second fragment from the obfuscated versions (from 15 to 45 seconds). For each pair, we compute matching scores using both Shazam (OSS) and our Topological Audio ID algorithm. The classification task is performed using the learned thresholds $\kappa_{\mathrm{Sh}}$ and $\kappa_{\mathrm{Top}}$ as discussed in \cref{sec:spotify_results} (refer to \cref{tbl:scores_grosser} for the scores).

While the open-source implementation of Shazam fails to recognize any of the songs, our algorithm successfully identifies 6 out of the 8 songs, with acceptable errors for the remaining 2 cases. It is important to note that this dataset was specifically designed in 2016 to challenge the audio ID systems of platforms like YouTube, Soundcloud, and Shazam at that time. The open-source version of Shazam primarily implements the algorithm described in \cite{wang_industrial-strength_2003}. However, the current version 15.35.0 (5369:f) of Shazam Mobile App \cite{shazam_2023} is now capable of recognizing four of the obfuscated songs from the 30-second snippets (indicated as $(*)$ in \cref{tbl:scores_grosser}). Nevertheless, out of the 6 songs correctly identified by our algorithm, half of them are still unrecognizable by the current version of Shazam \cite{shazam_2023}.

\begin{table}[htb!]
	\centering
	\begin{tabular}{|l|r|l|c|l|}
	    \hline
		\textbf{Song} &\multicolumn{2}{c|}{\textbf{Shazam (OSS)}}&\multicolumn{2}{c|}{\textbf{Top. Audio ID}} \\
        &Count&Prediction&Error&Prediction\\
		\hline
		Smells Like Teen Spirit
        &6&\textcolor{red}{Negative}
        &0.2244  &\textbf{Positive}\\
		Get Lucky 
        &12&Negative ${(*)}$
        & 0.0000 &Positive\\
		Giant Steps
        &8&\textcolor{red}{Negative}
        & 0.1117&\textbf{Positive}\\
		Stairway to Heaven
        &5&Negative ${(*)}$
        &0.0557 &Positive\\
		Headlines
        &5&Negative ${(*)}$ 
        &0.0838 &Positive\\
        Blue in Green
        &2&\textcolor{red}{Negative}
        &0.1989 &\textbf{Positive}\\
        You’re Gonna Leave
        &3&\textcolor{red}{Negative}
        &0.3677 &\textcolor{red}{Negative} \\
        Blue Ocean
        &2& \textbf{Negative ${(*)}$}
        &0.3416&\textcolor{red}{Negative}\\
		\hline
	\end{tabular}
\caption{\textbf{Classification results for Grosser's dataset.} The performace of Shazam (OSS) and our Topological Audio ID algorithm for the 8 positive positive pairs constructed from Grosser's dataset \cite{grosser_music_obfuscator}.
The columns \textit{Error} and \textit{Count} gives the outputs of each algorithm for every pair. The prediction is performed according to the thresholds $\kappa_{\mathrm{Top}}=0.2521$ and $\kappa_{Sh}=78$ learned from the training step with the Million Song dataset. That is, if the error is less than $\kappa_{\mathrm{Top}}$ (resp. count is greater than $\kappa_{\mathrm{Sh}}$), then the pair is classified as positive, otherwise is negative. The asterisks mean that the obfuscated snippet is recognized by Shazam Mobile App 2023 \cite{shazam_2023}. In red, wrong classifications, in bold the best performance.}
\label{tbl:scores_grosser}
\end{table}

\section{Discussion and future work}
\label{sec:discussion}

We presented a topological audio ID method for analyzing audio tracks. Our approach involves a fingerprinting method based on the persistent homology of spectral representations of audio signals, along with a tailored identification algorithm for robust content-based comparisons.
We tested our method using a range of obfuscated pairs of tracks and compared the results with the leading algorithm. The performance is comparable in rigid obfuscations for a reasonable range of distortion degrees. Notably, our algorithm performed better in handling topological obfuscations, in contrast to Shazam, which showed poor results.
Moreover, our algorithm achieved the correct identification of most of the challenging examples posed by Grosser \cite{grosser_music_obfuscator} (unrecognized by Shazam or the content ID algorithms on YouTube and Soundcloud in 2016). These results highlight the significance of content-invariance captured by persistent homology in addressing complex audio identification tasks, improving the reliability and robustness of identification algorithms.

In practical scenarios, it is often necessary to compare short fragments of unknown tracks with a database of well-identified complete audio tracks. For example, when dealing with songs, it is common to compare snippets of 20 or 30 seconds with a database of full songs, which typically have an average duration of 3 minutes. While our matching algorithm is initially designed for comparing pairs of tracks with similar lengths (such as 30 seconds versus 60 seconds), it can also be applied to a general case by subdividing the stored database into fixed-length blocks, (with a duration of, for instance, $T=60$ seconds). This approach ensures reliable matchings even when dealing with different track lengths.

Another possible challenge is the computational complexity of topological fingerprinting comparisons, that may be higher compared to the benchmark algorithm. To address this, a generalization for the \textit{$1$ vs $N$} problem should incorporate the use of hashing functions and efficient methods to explore the fingerprint database. A possible strategy involves replacing Betti curves with hash functions for persistence diagrams, as proposed by \cite{fasy_approximate_2018}. The generalization of our algorithm to the \textit{$1$ vs $N$} case is the subject of future work. 

\appendix
\section{Algebraic topology of image data}\label{appendix:cubical_complexes}
\subsection{Cubical complexes}
A \textit{cube} (or \textit{cubical cell}) $Q = I_1 \times \ldots \times I_d$ is a product of elementary intervals $I_1,\ldots, I_d$ of the form $\lbrack a,a \rbrack,\, \lbrack a, a+1 \rbrack,\, a\in\Z$. We say that
\begin{itemize}
	\item $d$ is the embedding number of $Q$,
	\item $\dim(Q) = \left\vert\{l \mid I_l\text{ is not degenerate}\}\right\vert$ is the dimension of $Q$,
	\item $Q$ is called a vertex if $\dim(Q)= 0$.
\end{itemize}
Cubes have geometric faces. A cube $Q'$ is said to be a \textit{face} of a cube $Q$ if $Q'\subset Q$. Moreover, if $\dim(Q_2) = \dim(Q)-1$, $Q'$ is a proper face of $Q$.
\begin{definition}
	Let $K$ be a collection of cubes of the same embedding dimension. Then, $K$ is a cubical complex if
	\begin{itemize}
		\item for any cube $Q \in K$, its faces are also in $K$,
		\item for all cubes $Q_1,Q_2 \in K$, the intersection $Q_1\cap Q_2 \in K$ is either empty or a face of $Q_1$ and $Q_2$,
	\end{itemize}
\label{def:cubical_complex}
\end{definition}

\begin{figure}[htb!]
	\centering
	\hfill
	\begin{tikzpicture}[scale=1.0, baseline=-5.5ex, transform shape,
		general/.style={draw,inner sep=0pt,minimum size=3pt},
		dark/.style={circle, fill=black!80},
		arrow/.style={->,>=stealth, shorten >=0.15cm},
		dCurrent/.style={regular polygon, regular polygon sides=3, fill=red!40},
		every label/.append style={font=\tiny}]
		
		\node[general, dark] (00) at (0,0) {};
		\node[general, dark] (01) at (0,-1) {};
		\node[general, dark] (02) at (0,-2) {};
		\node[general, dark] (10) at (1,0) {};
		\node[general, dark] (12) at (1,-2) {};
		\node[general, dark] (20) at (2,0) {};
		\node[general, dark] (21) at (2,-1) {};
		\node[general, dark] (22) at (2,-2) {};
		\node[general, dark] (11) at (1,-1) {};
		\draw (00) -- (01) -- (02) -- (12) -- (22) -- (21);
		\draw (00) -- (10) --coordinate[midway](0120) (20) --coordinate[midway](2021) (21);
		\draw (01) -- (11) -- (12);
		\draw (10) -- (11);
		
		\node[fill = gray!20, fit={(0,-1) (1,-2)}, inner sep=-1pt] (A) {};
		
		\node[align=center, text width=3cm] (vertex annotation) at (0.2, 1.5) {0-cube\\(Vertex)};
		\draw[arrow] (vertex annotation) -- (00);
		\draw[arrow] (vertex annotation) -- (10);
		
		\node[align=center, text width=3cm] (edge annotation) at (2.8, 1.3) {1-cube\\(Edge)};
		\draw[arrow] (edge annotation) -- (0120);
		\draw[arrow] (edge annotation) -- (2021);
		
		\node[align=center, text width=1.3cm] (square annotation) at (3.3, -1.4) {2-cube};
		\draw[arrow, shorten >=-0.2cm] (square annotation) -- (A);
		\end{tikzpicture}
		\hfill
		\begin{tikzpicture}[scale=1.0, baseline=-5.5ex, transform shape,
		general/.style={draw,inner sep=0pt,minimum size=3pt},
		dark/.style={circle, fill=black!80},
		dUp/.style={circle, fill=white!40},
		arrow/.style={->,>=stealth, shorten >=0.2cm},
		dCurrent/.style={regular polygon, regular polygon sides=3, fill=red!40},
		every label/.append style={font=\tiny}]
		
		\node[general, dark] (00) at (0,0) {};
		\node[general, dark] (01) at (0,-1) {};
		\node[general, dark] (02) at (0,-2) {};
		\node[general, dark] (10) at (1,0) {};
		\node[general, dUp] (11) at (1,-1) {};
		\node[general, dark] (12) at (1,-2) {};
		\node[general, dark] (20) at (2,0) {};
		\node[general, dark] (21) at (2,-1) {};
		\node[general, dark] (22) at (2,-2) {};
		\draw (00) -- (01) -- (02) -- (12) -- (22) -- (21);
		\draw (00) -- (10) -- (20) -- (21);
		\draw (01) -- (11);
		
		\node[align=center, text width=4cm] (vertex annotation) at (0.2, 1.1) {Vertex not in the complex};
		\draw[arrow] (vertex annotation) -- (11);
		\end{tikzpicture}
		\hfill
	\caption{\textbf{Cubical complexes.} On the left, an example of a collection of cubes of different dimensions, where two 0-cubes, two 1-cubes and one 2-cube are annotated. While the cubes are of different dimensions (0, 1 or 2), they share the embedding dimension 2. This collection is a cubical complex. On the right, a collection of cubes that is not a cubical complex. The annotated vertex is not in the collection, but the 1-cube incident to it is, violating the first property from \cref{def:cubical_complex}.}
	\label{fig:cubical_complex_example}
\end{figure}
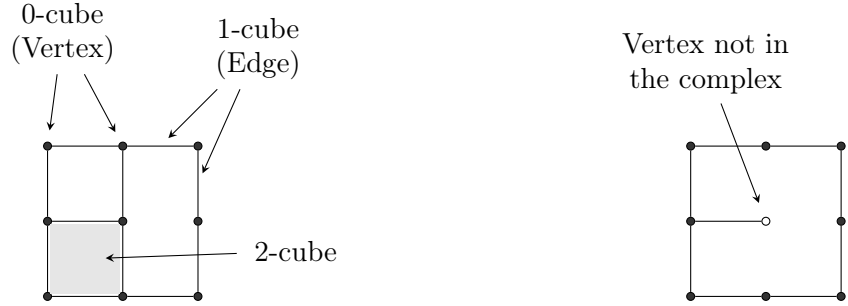
 
\begin{figure}[htb!]
	\input{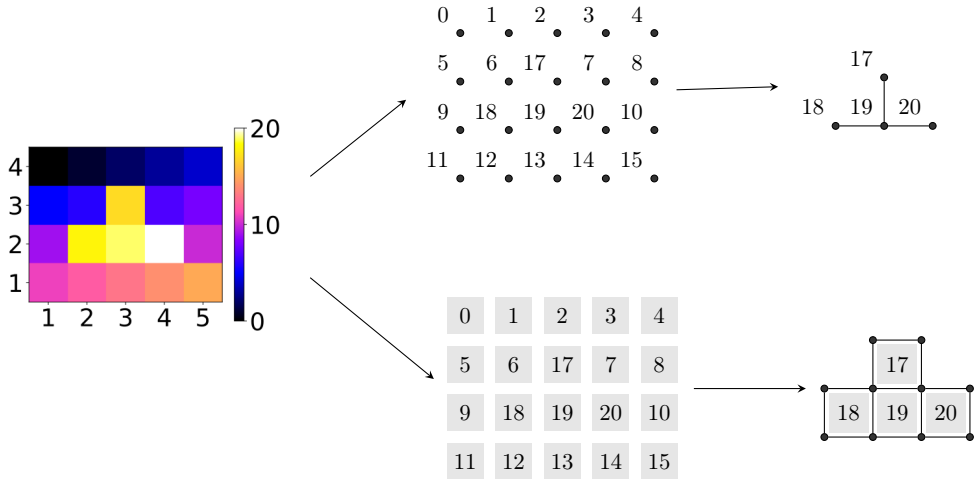}
	\hfill
	\caption{\textbf{V and T-constructions.}
		On the left, a grayscale image. In the vertex construction, at the top, each pixel from the image is a 0-cube, while in the top-cell construction, at the bottom, it is a 2-cube. The top-right figure illustrates the super-level set $K^{17}$. The 1 and 2 cubes were added to the 0 cubes on the figure in the centre and assigned filtration values according to the upper star co-filtration construction. An analogous process is shown on the bottom figure, where 0 and 1 cubes were added.
	}
	\label{fig:vertex_topcell_construction}
\end{figure}

There are two concurring ways of representing an image as a cubical complex \cite{garin_duality_2020}: the T-construction \cite{zeppelzauer_topological_2016} and the V-construction \cite{robins_percolating_2016}. In the former, each pixel corresponds to a 2-cell in the complex, while in the latter each pixel is a vertex, and a higher dimensional structure is built on it. The difference between the two constructions is illustrated in \cref{fig:vertex_topcell_construction}. 

A filtered (cubical) complex $K_*$ is a cubical complex $K$ together with a nested sequence of subcomplexes, that, a sequence of complexes $\emptyset=K_0 \subseteq K_1 \dots\subseteq K_m =K$.
Notice that the values of pixels in the image give us an $\R$-valued function on the vertices, which we will call a filter function. There are two ways natural ways to extend it to a filtration (or co-filtration).
\begin{definition}
	Let $f:V(K)\rightarrow \R$ be a function defined on the vertices $V(K)$ of a cubical complex $K$. The lower-star filtration associated to $f$ is $K_f = (K_s)_{s\in\R}$, where
	\begin{equation}
	K_s = \{Q \subset K \mid f(v) \leq s,\, \forall v \in V(Q)\}.
	\label{eq:lower_star}
	\end{equation}
	\label{def:lower_star}
\end{definition}
The upper star co-filtration $(K^s)_{s\in\R}$ is defined analogously to \cref{def:lower_star}, but reversing the order in \cref{eq:lower_star},
\begin{equation}
K^s = \{Q \subset K \mid f(v) \geq s,\, \forall v \in V(Q)\}.
\label{eq:upper_star}
\end{equation}

\subsection{Homology and persistent homology} \label{appendix:homology} Let $K$ be a cubical complex.
For $d\in \mathbb{Z}$, let $K_d$ be the set of $d$-dimensional cells in $K$. The $\mathbb{F}_2$-vector space $C_d(K)$ is the \textit{chain group} of $K$ in degree $d$, which is freely generated by $K_d$. The chain groups are connected by linear boundary maps $\partial_d : C_d(K) \to C_{d-1}(K)$, which map a cell to the sum of its faces of codimension 1 and are extended linearly to all of $C_d(K)$ (see \cite[Sect. 2.2.3]{kaczynski_computational_2011} for more details). The \textit{cubical chain complex} $C_*(K)$ is defined as the pair $(\{C_d(K)\}_{d\in \mathbb{Z}}, \{\partial_d\}_{d\in \mathbb{Z}})$. Denote the subspace of \textit{cycles} by $Z_d(K) = \ker \partial_d$ and the subspace of \textit{boundaries} by $B_d(K) = \mathrm{im} \partial_{d+1}$. Since $\partial_{d-1} \circ \partial_d = 0$, every boundary is a cycle, and then the \textit{homology group} of $K$ in degree $d$ is defined as the quotient space $H_d(K) := Z_d(K)/B_d(K)$. Note that the homology groups retain the structure of  $\mathbb{F}_2$-vector space, and their dimension $\beta_d(K) = \dim{\mathbb{F}_2} (H_d(K))$ represents the $d$-th \textit{Betti number} of $K$.

Persistent homology \cite{zomorodian_computing_2005} is an extension of homology for filtrations (or co-filtrations) of complexes. Consider an ordered set $(S, \leq)$ and a filtration of complexes $(K^s)_{s\in\mathcal{S}}$\footnote{An equivalent theory can be developed for co-filtrations of complexes.}. The inclusion morphisms $K^s \hookrightarrow K^{t}$ for $s\leq t$ induce morphisms between homology groups $\iota_{s,t}: H_k(K^s) \rightarrow H_k(K^t)$. The pair $\left(\{H_k(K^t)\}_{t\in S}, \{\iota_{s,t}\}_{\substack{s\leq t\\ s,t \in S}}\right)$ is a \textit{persistence module} of finite type.
The $\mathbb{F}_2[t]$-graded module $M:=\bigoplus_{t\in S} H_k(K^t)$ is isomorphic to a direct sum
of interval modules $M \simeq \bigoplus_{B(M)} Q(I)$. Here $B(M)$ denotes the \textit{barcode} of $M$, which is given by a multiset of intervals, and the  \textit{interval module} $Q(I)$ for an interval $I\subseteq \mathbb{R}$ is defined as
\[Q(I_s)=\begin{cases}\mathbb{F}_2 & \text{if }s\in I\\
0&\text{otherwise,}\end{cases}\] with maps defined as the identity inside the interval. For a complete exposition of these results, see \cite{crawley2015decomposition, Botnan2018DecompositionOP}.

\section*{Acknowledgments}
We would like to thank Prof. Kathryn Hess for facilitating the student internship with HAH and thank the Mathematical Institute at Oxford for hosting WR. 
 We thank Martin Gould, Johnny Hunter,  Shahar Elisha and Linden Vongsathorn for their help and continuous support. HAH acknowledges funding from a Royal Society University Research Fellowship. HAH and XF are partly funded by EPSRC EP/R018472/1 supporting the Centre for Topological Data Analysis.

\bibliographystyle{siamplain}
\bibliography{references}
\end{document}